\documentstyle[epsf,rotate,german]{mn}
\newif\ifAMStwofonts
\AMStwofontstrue

\def\O{$\Omega$}
\input wasyfont

\def\ar{{$\rightarrow$}}

\def\D{{$\Delta$}}
\def\De{{$\Delta$ }}
\def\cx{\mbox{\raisebox{.4mm}{\tiny{$\supset$\hspace{-0.6mm}$\subset$}}}}
\def\cxs{\mbox{\raisebox{.4mm}{\tiny{$\supset$\hspace{-0.6mm}$\subset$ }}}}
\def\emp{\mbox{\raisebox{.4mm}
{\tiny{$\rangle$\hspace{-.6mm}$\bigcirc$\hspace{-.55mm}$\langle$}}}}

\ifoldfss
  \ifCUPmtlplainloaded \else
    \NewTextAlphabet{textbfit} {cmbxti10} {}
    \NewTextAlphabet{textbfss} {cmssbx10} {}
    \NewMathAlphabet{mathbfit} {cmbxti10} {} 
    \NewMathAlphabet{mathbfss} {cmssbx10} {} 
  \fi
  \ifAMStwofonts
    \ifCUPmtlplainloaded \else
      \NewSymbolFont{upmath} {eurm10}
      \NewSymbolFont{AMSa} {msam10}
      \NewMathSymbol{\upi}     {0}{upmath}{19}
      \NewMathSymbol{\umu}     {0}{upmath}{16}
      \NewMathSymbol{\upartial}{0}{upmath}{40}
      \NewMathSymbol{\leqslant}{3}{AMSa}{36}
      \NewMathSymbol{\geqslant}{3}{AMSa}{3E}

    \fi
  \fi
\fi 

\ifnfssone
  \newmathalphabet{\mathit}
  \addtoversion{normal}{\mathit}{cmr}{m}{it}
  \addtoversion{bold}{\mathit}{cmr}{bx}{it}
  \newmathalphabet{\mathbfit} 
  \addtoversion{normal}{\mathbfit}{cmr}{bx}{it}
  \addtoversion{bold}{\mathbfit}{cmr}{bx}{it}
  \newmathalphabet{\mathbfss} 
  \addtoversion{normal}{\mathbfss}{cmss}{bx}{n}
  \addtoversion{bold}{\mathbfss}{cmss}{bx}{n}
  \ifAMStwofonts
    \ifCUPmtlplainloaded \else
      \UseAMStwoboldmath
      \makeatletter
      \new@mathgroup\upmath@group
      \define@mathgroup\mv@normal\upmath@group{eur}{m}{n}
      \define@mathgroup\mv@bold\upmath@group{eur}{b}{n}
      \edef\UPM{\hexnumber\upmath@group}
      \new@mathgroup\amsa@group
      \define@mathgroup\mv@normal\amsa@group{msa}{m}{n}
      \define@mathgroup\mv@bold\amsa@group{msa}{m}{n}
      \edef\AMSa{\hexnumber\amsa@group}
      \makeatother
      \mathchardef\upi="0\UPM19
      \mathchardef\umu="0\UPM16
      \mathchardef\upartial="0\UPM40
      \mathchardef\leqslant="3\AMSa36
      \mathchardef\geqslant="3\AMSa3E
    \fi
  \fi
\fi 

\ifnfsstwo
  \DeclareMathAlphabet{\mathbfit}{OT1}{cmr}{bx}{it}
  \SetMathAlphabet\mathbfit{bold}{OT1}{cmr}{bx}{it}
  \DeclareMathAlphabet{\mathbfss}{OT1}{cmss}{bx}{n}
  \SetMathAlphabet\mathbfss{bold}{OT1}{cmss}{bx}{n}
  \ifAMStwofonts
    \ifCUPmtlplainloaded \else
      \DeclareSymbolFont{UPM}{U}{eur}{m}{n}
      \SetSymbolFont{UPM}{bold}{U}{eur}{b}{n}
      \DeclareSymbolFont{AMSa}{U}{msa}{m}{n}
      \DeclareMathSymbol{\upi}{0}{UPM}{"19}
      \DeclareMathSymbol{\umu}{0}{UPM}{"16}
      \DeclareMathSymbol{\upartial}{0}{UPM}{"40}
      \DeclareMathSymbol{\leqslant}{3}{AMSa}{"36}
      \DeclareMathSymbol{\geqslant}{3}{AMSa}{"3E}
    \fi
  \fi
\fi 

\ifCUPmtlplainloaded \else
  \ifAMStwofonts \else 
    \def\upi{\pi}
    \def\umu{\mu}
    \def\upartial{\partial}
  \fi
\fi

\title{Orbital dynamics of three-dimensional bars: \\
III. Boxy/Peanut edge-on profiles}

\author[P.A.~Patsis et al.]
{P.A.~Patsis,$^1$ Ch.~Skokos,$^{1,2}$ E.~Athanassoula$^3$\\
$^1$Research Center of Astronomy, Academy of Athens, Anagnostopoulou 14,
  GR-10673 Athens, Greece\\
$^2$ Division of Applied Analysis, Department of Mathematics and Center for
Research and Application of Nonlinear Systems (CRANS),\\ University of Patras,
GR-26500, Patras, Greece\\
$^3$Observatoire de Marseille, 2 Place Le Verrier, F-13248 Marseille, Cedex 4,
  France}
\date{Accepted ....
      Received ....
      in original form ....}
\pagerange{\pageref{firstpage}--\pageref{lastpage}}
\pubyear{2001}

\begin{document}

\maketitle

\label{firstpage}

\begin{abstract}
  We present families, and sets of families, of periodic orbits that provide
  building blocks for boxy and peanut (hereafter b/p) edge-on profiles. We
  find cases where the b/p profile is confined to the central parts of the
  model and cases where a major fraction of the bar participates in this
  morphology. A b/p feature can be built either by 3D families associated with
  3D bifurcations of the x1 family, or, in some models, even by families
  related with the $z$-axis orbits and existing over large energy intervals.
  The {\sf `X'} feature observed inside the boxy bulges of several edge-on
  galaxies can be attributed to the peaks of successive x1v1 orbits (Skokos et
  al. 2002a, hereafter paper I), provided their stability allows it. 
  However in general, the x1v1 family has to overcome the obstacle of
  a S\ar\D\ar S transition in order to support the structure of a b/p feature.
  Other families that can be the backbones of b/p features are x1v4 and
  z3.1s.
  The morphology and the size of the boxy
  or peanut-shaped structures we find in our models is determined by the
  presence and stability of the families that support b/p features.  The
  present study favours the idea that the observed edge-on profiles are the
  imprints of families of periodic orbits that can be found in appropriately
  chosen Hamiltonian systems, describing the potential of the bar.
\end{abstract}


\section{Introduction}

Disk galaxies, when observed edge-on, often exhibit a box- or peanut-like 
structure. Since this is confined to the inner
parts of the galaxy, and since it extends in the vertical direction
outside the disk, this structure has been called a boxy or
peanut bulge. Yet there are many ways in which it differs from
ordinary bulges.
The b/p structures have their maximum thickness not at the center of the
galaxy, like in usual $R^{1/4}$ spheroids, but `at two points symmetrically
spaced on either side of the center' \cite{bb59}. Another difference from
$R^{1/4}$-bulges and ellipticals is that b/p structures rotate
`cylindrically', i.e. their observed rotation is independent of the height
above the plane, which bulges and ellipticals do not (e.g.
Kormendy \& Illingworth 1982). There is one more characteristic of b/p
features related to their kinematics. Kuijken \& Merrifield
(1995) and Bureau \& Freeman (1999) have shown
that there are important differences between the position velocity
diagrams of b/p structures and those of bulges. These have been used
by Bureau \& Athanassoula (1999) and Athanassoula \& Bureau (1999) to
develop diagnostics to detect the presence and orientation of a bar
in edge-on disk galaxies. The method relies on the presence of x2 orbits in
the bars of the galaxies. Seen these differences, we will avoid
calling the b/p features bulges, unless we are refering to particular
observations.

Recent statistical studies (L"utticke, Dettmar \& Pohlen 2000a) using 1350
galaxies from the RC3, show that 45\% of the profiles of edge-on disc galaxies
are box- or peanut-shaped. Observational studies (e.g. Bureau \& Freeman 1999,
L"utticke, Dettmar \& Pohlen 2000b) associate the b/p structure with the
presence of a bar. L"utticke et al. (2000b) classify the bulges according to
their boxiness and conclude that galaxies with a prominent b/p shape have a
large BAL/BUL ratio, where BAL is the projected bar length, and BUL the bulge
length. Isolating photometrically the b/p structure from the bulge, they
measure the ratio of the projected bar length to the length of the b/p
structure (BAL/BPL). Unfortunately this is done only for six galaxies,
and gives an average value of $2.7\pm0.4$. This ratio indicates a
structure confined close to the center of the galaxy.

Bars and edge-on b/p morphology are linked in all the above mentioned
papers, as well as in many others. However, the percentage of the bar
which takes part in the b/p structure,
i.e. whether we have a b/p feature on top of the bar, or whether we have a
b/p-shaped bar in total, is an open question.  The morphological differences
encountered among the various b/p features remains also to be explained.

In a few cases (e.g. IC~4767, Hickson 87a) an `{\sf X}'-shaped structure is
found to be embedded in a boxy structure. It differs from the usual peanut in
that the branches of the {\sf `X'} feature resemble segments of nearly
straight lines that give the impression of intersecting each other. On the
contrary the classical peanuts have typically much rounder isophotes (see e.g.
Shaw 1993) and, even in cases where these isophotes come very close to the
equatorial plane of the galaxy at the center, the visual impression is better
described by the symbol `\cx' than by an `{\sf X}' central
morphology\footnote{nice examples can be found in the web page of L.
  Kuchinski at \\{\em
    http://www.astronomy.ohio-state.edu/\~{}lek/galx.html}}. L"utticke, 
Dettmar \& Pohlen (2000c), studying a sample of b/p galaxies including cases
with {\sf `X'} features, estimated the angle between one branch of the {\sf
  `X'} and the major axis to be around $40 \degr \pm 10 \degr$. Studies of
individual galaxies give for this angle values ranging from $22\degr$ for
IC~4767 \cite{whib88} to $45\degr$ for NGC~128 \cite{do99}. Pfenniger \&
Friedli (1991), taking as an example the case of IC~4767, claim that this
feature is an optical illusion obtained when one uses a particular
look-up-table for viewing the image.  This view, however, is not generally
shared. Thus Mihos, Walker, Hernquist et al.  (1995) used the process of
`unsharp masking' to enhance the {\sf `X'} embedded in the bulge of the galaxy
Hickson 87a in order to compare this morphology with their model. In any case
one can speak about characteristic kinks of the isophotes in edge-on profiles
of a few galaxies and of the corresponding isodensities in snapshots of some
$N$-body simulations \cite{am01}, which are aligned in such a way as to
describe an {\sf `X'}-shaped feature.

There have been two approaches for explaining edge-on b/p profiles.
The first invokes internal reasons, like disk or orbital
instabilities, and the second one external
reasons, like encounters with companions, soft merging etc.

Combes \& Sanders (1981) were the first to reproduce a b/p profile in $N$-body
simulations of barred galaxies. Such structures were since then found
in many other
simulations (e.g. Combes, Debasch, Friedli et al. 1990; Raha, Sellwood, James
et al. 1991; Athanassoula \& Misiriotis 2002) and are now considered a
standard development in bar-unstable disc simulations. Pfenniger (1984a, 1985)
associated the b/p morphology with the instability of the x1 family at the 4:1
vertical resonance. Later, Combes et al. (1990) suggested that the b/p shapes
are due to the vertical 2:1 resonance, stressing the importance of having
$\Omega_b = \Omega - \kappa /2 = \Omega - \nu /2$. This mechanism invokes a
conjunction of the two resonances, i.e. the radial 2:1 (radial ILR) and the
vertical 2:1 (v-ILR), and relates it to the appearance of a b/p
edge-on morphology.
The 3D families introduced at higher order resonances exist typically over
smaller energy intervals and thus are less probable to be populated by orbits
to build the box, although in principle they could be used as well. In such a
case of course, they will support a thin morphological feature extending to
large distances from the center, i.e. close to
corotation. The 2:1 vertical resonance is proposed as explanation of boxy
structures also by Pfenniger \& Friedli (1991), who speak about thick b/p
bars. We mention that b/p features have also been found in edge-on profiles of
orbital models of {\em normal} spiral galaxies with thick, 3D spirals embedded
in discs \cite{pg96}. Berentzen, Heller, Shlosman et al. (1998) have shown
that a peanut shape may disappear when there is substantial gas inflow
to the center of the galaxy.

Building the peanut by accretion of material from satellite galaxies
has been proposed initially by Binney \& Petrou (1985), while Mihos et
al. (1995) describe an encounter which produces the {\sf `X'} feature
in the galaxy Hickson 87a. However, even in the cases where a
companion is involved, the families of orbits that trap the infalling
gas have to be studied.

In the present paper we investigate the vertical structure of bars using
orbital theory. We do not construct self-consistent models, but we explore
changes that occur when the main parameters of the system vary. We combine
families of periodic orbits found in the models of paper I and in the models
of Skokos, Patsis and Athanassoula (2002b - hereafter paper II), in order to
build b/p features in the edge-on profiles. We also compare the geometry and
the dimensions of the resulting systems with the corresponding features of
edge-on galaxies. Speaking about a `bulge' in the profiles of our models, we
refer to a central enhancement of the density due to 3D orbits in our total
potential. As we will see, the families of orbits we use are mainly 3D
bifurcations of the planar x1 orbits, i.e. related to the families of the
x1-tree that make the 3D bar in our models.

The layout of this paper is as follows: In Section~2 we describe the
method we used to construct the vertical profiles of the families in
the models and in Section~3 and 4 we describe the properties of these
profiles and the effect of combining several families together on the
morphology of the models. In Section~5 we discuss our results and
compare them with observations found in the literature, and finally in
Section~6 we enumerate our conclusions.

\section{Method}
\subsection{The families of periodic orbits}
Our general model consists of a Miyamoto disc, a Plummer bulge and a
Ferrers bar. The total mass of the Miyamoto disc is $M_D$, and its
horizontal and vertical scalelengths are $A$ and $B$ respectively.
We have taken in all models $A=3$ and $B=1$.  The total mass of the
Plummer bulge is $M_S$ and its scalelength $\epsilon_s$. The mass of
the Ferrers bar component is indicated by $M_B$ and its semi-axes by
$a$, $b$ and $c$, where $a:b:c$ = $6:1.5:0.6$. The masses of the three
components satisfy $G(M_D+M_S+M_B)=1$, where $G$ is the gravitational
constant. The unit length is taken as 1~kpc, the time unit 1~Myr and
the mass unit 2$\times$10$^{11}$M$_{\odot}$.  Details can be found in
Section 3.1 of Paper I, while the values of the parameters
characterizing each particular model are given in Tables in the
corresponding sections of the present paper.

In Papers I and II we calculated the families of periodic orbits which
are the building blocks of 3D bars.  Paper I presents in a fiducial
case the x1-tree and the typical orbital behaviour of 3D bars in
detail, while paper II describes the differences observed in the
orbital structure when the main parameters of the model vary. These
papers also give the morphology of the orbits of each family. For the
purpose of the present paper we are interested in the edge-on (side-on
and end-on\footnote{in the side-on view the line of sight is along the 
minor axis of the bar and in the end-on view it is along the major axis}) 
structures supported by the families, when one considers
them as dynamical blocks. In this section we review briefly the
properties of the main families we will discuss below.  Throughout the
paper the bar major axis lies along the $y$ axis and the axis of rotation
is the $z$ axis.

\begin{description}
\item {\em x1v1}. In most of the models this is the first stable
  simple-periodic vertical bifurcation of x1, i.e. it bifurcates at the lowest
  energy at which the stability index associated with the vertical
  perturbations becomes $\lid -2$. It is related to the 2:1 vertical resonance
  and both its $(x,z)$ and $(y,z)$ projections can be described as `frowns'
  (`$\frown$') and `smiles' (`$\smile$') . Several papers associate this
  family with the b/p bulges by
  combining its two branches which are symmetric with respect to the 
  equatorial plane
  (Combes et al. 1990, Pfenniger \& Friedli 1991). In many cases this family
  has a complex unstable part close to the energy at which it bifurcates from
  x1 (for an extensive description of complex
  instability in galactic potentials see e.g. Zachilas 1993).
\item {\em x1v3}. It is the stable 3D bifurcation of the models at the
  vertical 3:1 resonance. It is one of the standard building blocks of the
  3D bar. Its $(x,z)$ projection has a tilde-like shape
  `({\large $\sim$})', while the $(y,z)$ one is a kind of a
  `o\hspace{-0.2mm}o\hspace{-0.2mm}o' figure.
\item {\em x1v4}. This family is bifurcated as unstable from x1, also
  at the vertical 3:1 resonance region, but
  in many models it has large stable parts for larger energies. Thus
  frequently it is an important family of the system. Its $(x,z)$ projection
  resembles the $(y,z)$ projection of x1v3, i.e. it is a
  `o\hspace{-0.2mm}o\hspace{-0.2mm}o' figure, while the $(y,z)$ one has a
  tilde-like shape.
\item {\em x1v5}. It is bifurcated at the vertical
  4:1 resonance. Both  its edge-on projections can be morphologically described
  with the letter  `{\sf w}', and it was the first family proposed for
  building the peanut (Pfenniger 1984a, 1985). In some models we have two
  x1v5-like families.
\item {\em z3.1s}. This family is important in the model
  which lacks an explicit Plummer sphere bulge component in the
  potential. Morphologically it resembles the x1v4 family, but it is not
  related to the x1-tree.
\item {\em x2-like 3D families}. Finally, in a couple of cases, we found
  stable 3D x2-like families, i.e. families whose $(x,y)$ projections
  have elliptical-like shapes
  elongated along the bar {\em minor} axis. They
  are
  bifurcations from the planar x2 orbits.
\end{description}
We note that  all families have counterparts which are symmetric with respect
to the principal planes and one can take into account all these orbits
in order to build a profile. The rest of the stable 3D families found in
papers I and II also contribute to the vertical profiles of the models,
but play a less important role. In Section 3, we refer to some of them
in the description of the profiles of each model.

\subsection{The weights of the  orbits}
In order to estimate the edge-on morphologies that can be supported we
use, as a tool, profiles built by sets of weighted periodic orbit. For this
we first calculate a set of stable periodic orbits and pick points
along each orbit at equal time steps. We also calculate the `mean
density' of each orbit, by calculating at each step the density of the
model at the point visited by the orbit, and then taking the mean of
these values. This `mean density' is used to weight the orbit, since 
it can be considered as a first approximation
for its importance. Thus the relative significance of the
stable families for the dynamics of our model can be estimated by weighting
the orbits by their `mean density'. This
is particularly important in regions where several families
coexist. Using a program specifically made for this task and based on
ESO-MIDAS routines, we construct an image (normalized over its total
intensity) for each calculated and weighted orbit. The selected
weighted orbits are then combined to build a profile. Darker parts of
the profile image correspond either to regions occupied by a single
family of relatively high importance, or to regions where many
different orbits coexist.

To give a visual impression of the morphology supported by the different
stable families we consider orbits from all families that could play a role.
These are families having stable parts for large energy intervals. An
exception are families born at the radial 3:1 resonance region, since as we
have seen in papers I and II, their role for the dynamics of the studied
models is only local. The selected orbits are equally spaced in their mean
radius. This step in mean radius was the same for all families in a model. By
over-plotting these orbits one gets a crude impression of the edge-on
structure of a stellar model, constructed on the basis of the selected
families.  Such figures are not density maps like those presented in the 2D
case in Contopoulos \& Grosb{\o}l (1988) and Patsis, Contopoulos \& Grosb{\o}l
(1991), since we deal here only with periodic orbits. The diagrams should
rather be considered as skeletons of supported structures.

The 3D families appear in pairs symmetric with respect to the $x$=0, $y$=0 and
$z$=0 planes. The $(x,y)$ projections of the orbits of most families are
identical to those of their symmetric pairs. That means that in most cases
there is no morphological difference in the face-on view if we add the
symmetric branches.  On the other hand, in several cases, it is necessary to
populate all symmetric branches to obtain symmetric edge-on profiles.

We present for all our models both the profiles of each family
separately and also combined profiles, where we consider all or some
of the families of the model. When we estimate the relative
importance of periodic orbits for the building of the vertical 
profile in a galaxy we have to take into account the fact that stable 
periodic orbits may reach large distances in the $z$-direction. This
means that stable orbits with high $\overline{|z|}$ values are
of lesser importance than the orbits remaining close to the equatorial plane
and enhancing the bar. For the sake of completeness, however, we included
these orbits in our profiles and we comment on their importance in each
individual case.

\section{Profiles of families}
In this section we present the edge-on profiles of the various
families in each model. In the present paper we again use model A1 as
a fiducial case. We describe the contribution of all its families
to the profiles, while for the rest of the models we give mainly the
differences of their orbital behaviour from what we find in model A1.
\subsection{Model A1}
\begin{figure*}
\epsfxsize=15.0cm \epsfbox{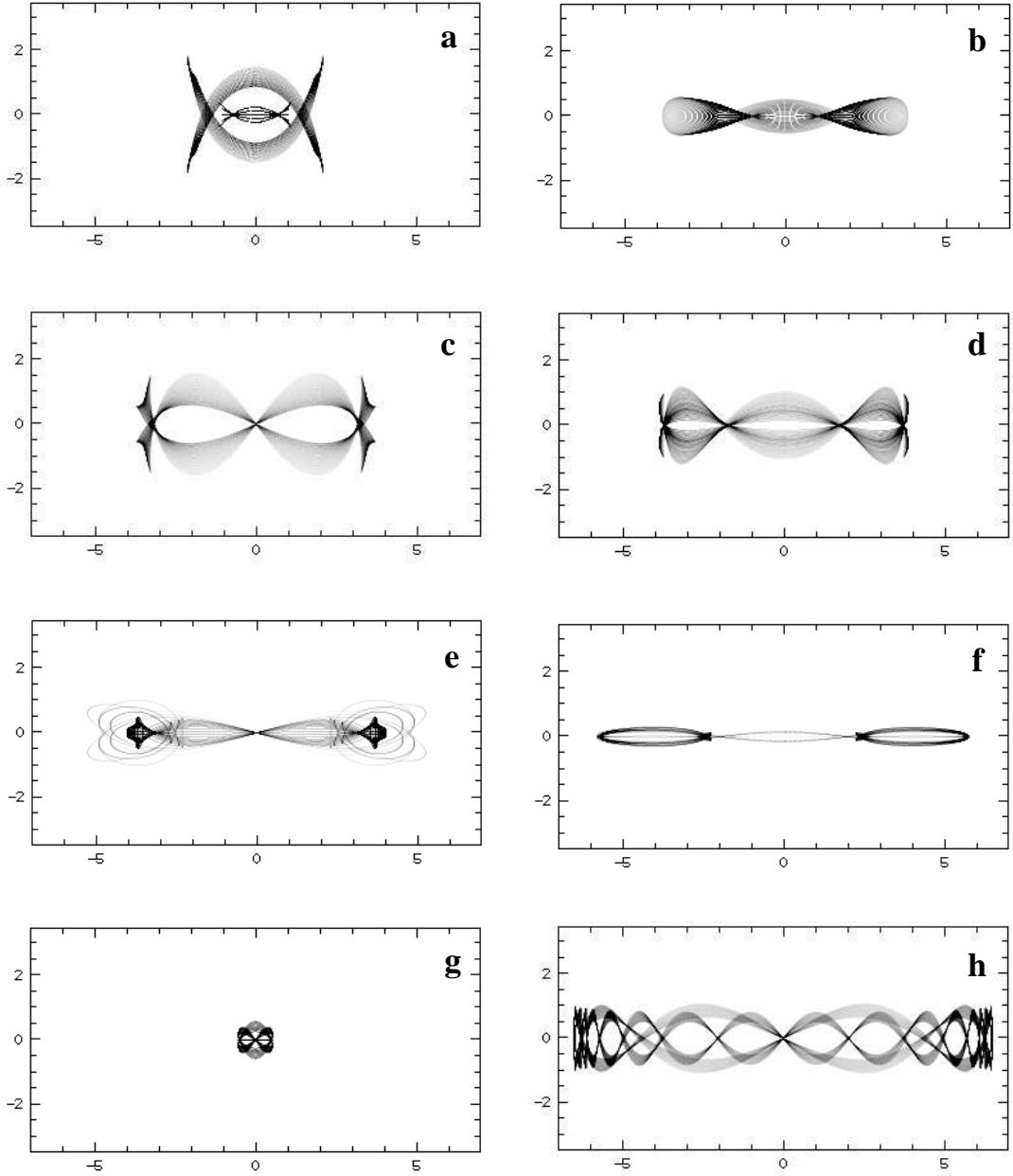}
\caption[]{The profiles of the 3D families of model A1 viewed side-on. Each
  panel corresponds to one 
family: (a) x1v1, (b) x1v3, (c) x1v4, (d) x1v5, (e) x1v7, (f) x1v9,
(g) x2mul2, and (h) ban3v1. In model A1 corotation is at 6.13. We use
a different contrast in each figure in order to bring out best the
morphology of the profile of each family. }
\label{A1yzall}
\end{figure*}

Model A1, introduced in paper~I, includes a typical Ferrers bar, and
has both a radial and a vertical 2:1 resonance.  In Fig.~\ref{A1yzall}
we give for this model the side-on profiles of all important 3D
families we found in paper I. All panels are profiles of individual
families. Throughout the paper we use a linear gray scale (i.e. a linear
look-up-table in MIDAS) to display images.

Visually we can separate the profiles in two classes. Those for which the
projections of their orbits pass through the projected center of the
system, i.e. the (0,0) point, and those for which the (0,0) point is at the
center of an empty region. To the first class belong the orbits of
the families x1v4 (Fig.~\ref{A1yzall}c), x1v7 (Fig.~\ref{A1yzall}e),
the family of the 3D x2-like orbits of multiplicity 2
(Fig.~\ref{A1yzall}g) and the 3D banana-like orbits of the family
ban3v1 (Fig.~\ref{A1yzall}h). Two of these families, x1v4 and x1v7, belong to
the 
x1-tree. Their common characteristic is that they are bifurcations of
x1 in $\dot{z}$, i.e. they have initial conditions
$(x_{0},\dot{x}_{0},z_{0},\dot{z}_{0}) = (a,0,0,b)$, with $a,b\in
\mathbf{R}$ and $a,b \neq 0$, while $y_{0}=0$ and $\dot{y}_{0}>0$. We
note that the structures they support 
are more `\cx'-like than {\sf `X'}-like features. In that respect these
families favour the presence of normal and not {\sf `X'}-shaped
peanuts.

The profiles that result from the superposition of stable orbits of the
families x1v1 (Fig.~\ref{A1yzall}a), x1v3 (Fig.~\ref{A1yzall}b), x1v5
(Fig.~\ref{A1yzall}d) and x1v9 (Fig.~\ref{A1yzall}f), belong to the
second class of profiles, which have an empty central region.

In order to quantify the morphologies supported in the various cases
we use the $B_L/O_{Ly}$ ratio. Here $B_L$ is the bar length and
$O_{Ly}$ is the orbital length of the family under
consideration. $B_L$ is estimated by considering the extent of all
stable orbits {\em supporting} the bar. These are in general families
of the x1-tree (Papers I \& II). We project them on the semi-major
axis, and the longest projection is $B_L$.  In some cases (see the
case of x1v7 family below) we give in addition a second value for the
$B_L/O_{Ly}$ ratio taking into account for the estimation of $B_L$
also orbits of a family which do not fully support the bar (e.g. they
have loops off the major axis, which extend in large distances). This
number can be less than 1. We estimate $O_{Ly}$, the orbital length,
again by projecting orbits on the semi-major axis. However, for
$O_{Ly}$, we take into account only the orbits of the family we
consider. The `bar length' characterizes a model globally, while the
`orbital length' characterizes the $(y,z)$ projection of a given
family in the model. Since we investigate the contribution of the
families to the boxy structures and to b/p features in general, these
ratios should be compared with the corresponding ratios given by
L"utticke et al. (2000b) for observations and Athanassoula \&
Misiriotis (2002) for $N$-body simulations. This will be done in
Section~5.

In particular we can remark about the profiles of model A1 the following:

In general, the families which have at the center an `\cx' morphology are good
candidates to build, or at least to support, the peanut feature. Family x1v4
(Fig.~\ref{A1yzall}c) has an `$\infty$'-type profile with two density
enhancements added to its sides. Tangents to the inner parts of this
`$\infty$' feature passing through the center have an angle to the major axis
of $\approx 22\degr$. The empty inner parts of the `$\infty$' feature reflect
the fact that this family is born as unstable and thus lacks members lying
almost on the equatorial plane, i.e. with low $\overline{|z|}$. The `bar
length', in model A1 reaches $r\approx 4.5$ (Patsis, Skokos, Athanassoula
2002b, hereafter paper IV) thus for the family x1v4 we have $B_L/O_{Ly} \approx
1.3$.

x1v7 (Fig.~\ref{A1yzall}e) also clearly harbours an `\cx' feature. The
angle to the major axis now is $\approx 12\degr$. A box is vaguely
defined, but now it is thin since the orbits remain close to the
equatorial plane.  The profile is better described as having a
`$\infty$' morphology with a low $\overline{|z|}$ and with two
characteristic local enhancements of the surface density at $|y|
\approx 3.8$ along the major axis. The $B_L/O_{Ly}$ ratio is $\approx
1.2$ if we consider as edges the enhancements, or $\approx 0.9$ if we
include the few outermost orbits as well. These outermost orbits are
4:1 rectangular-like orbits with loops at their four apocentra in
their $(x,y)$ projections. Their contribution to the density of the
bar and to its face-on morphology is discussed in Paper
IV. Nevertheless these outermost orbits are not considered as
contributing to the extent of the bar towards corotation.

Note that both x1v4 and x1v7 have local enhancements along the major axis,
which are symmetric with respect to the center and are manifestations of the
orbital character of the profile. If the edge-on profile of a galaxy is
determined by families like x1v4 and x1v7, then both the `\cx'
morphology and the local enhancements on the major axis should be 
observed (see Section 5 below).

The peanut that could be supported by the x2mul2 family is confined
to the central region of the disc (Fig.~\ref{A1yzall}g). This is
expected because the projections of the orbits of this family on the
equatorial plane are oriented along the minor axis of the main
bar. For this family we have $B_L/O_{Ly} \approx 9$. x2mul2 could be
useful to explain structures embedded in the very centers of
edge-on disc galaxies.

Finally the ban3v1 3D banana-like orbits favour in this projection the
presence of an `\cx' structure in the central region of a model.

Side-on profiles with empty regions around the central point (0,0) are
provided by the families x1v1 and x1v5, both proposed in the past as
peanut-building families (Combes et al. 1990, Pfenniger \& Friedli 1991;
Pfenniger 1984a, 1985). The x1v1 profile (Fig.~\ref{A1yzall}a) consists of two
parts, an inner and an outer one. This reflects the fact that the evolution of
the stability of family x1v1 follows a S\ar \D\ar S transition (paper I) and
thus its stable parts exist for two separated energy intervals. The gap is
formed because only stable orbits should be considered. Actually the missing
complex unstable part of the family would have provided very useful members
for building the boxy bulge, in the sense that this family now provides only
orbits with either very low or with relatively high $\overline{|z|}$. The
outer part has at the left and right sides `wings', which could bring in
the system pieces of an {\sf `X'} structure. These pieces, however, do not
continue towards the center. The central area separates the two pieces of the
{\sf `X'}. We will refer to this kind of morphology with the symbol `\emp'. We
note that the wings reach from the equatorial plane to their maximum height
very abruptly over a distance $\Delta y \approx 0.4$ and that the successive
weighted orbits overplotted form four sharp peaks.  The `orbital length'
reaches a distance from the center $r\approx 2.1$, so the $B_L/O_{Ly}$ ratio
for family x1v1 is about 2.1. If the outer part of the family, after the S\ar
\De transition, is not populated, then we have $B_L/O_{Ly} \approx 4.1$. In
this case, however, we would have orbits confined close to the center with
small deviations from the equatorial plane. Velocity dispersion will smooth
out such features and due to dust they will not be easily observed in edge-on
real galaxies.

Families x1v3 and x1v5, although bifurcated at different vertical resonances,
can give similar side-on profiles (cf. Fig.~\ref{A1yzall}b and d). This is a
good example to show that it is the superpositions of stable orbits of a
family, i.e. the vertical profile in a model, and not just the morphology of
single orbital representatives, that can decide about the supported
morphologies.  Since the `bar length' in model A1 is $\approx 4.5$, both
families have $B_L/O_{Ly} \approx 1.2$.

Family x1v9 remains always close to the equatorial plane. It does not directly
support the bar, since its $(x,y)$ projection consists of 4:1 rectangular like
orbits with big loops. Furthermore, it does not contribute much to specific
features, like boxes, observed in vertical profiles. Since it is not
considered as a family helping the bar to come closer to corotation, we do not
give in this case orbital ratios.

A common feature encountered in all four families in model A1 with a `\emp'
vertical morphology is that, independent of the vertical resonance at which
they are bifurcated, their side-on profiles are characterized by three local
maxima above the equatorial plane. This is a disadvantage
for using them for building peanuts, since b/p bulges have two such local
maxima. These families could nevertheless give to the model an overall boxy
side-on morphology, but only if we consider orbits trapped around the
periodic ones. This can be easier
done with the x1v1 family because its orbits reach higher $\overline{|z|}$
distances.

\begin{figure*}
\epsfxsize=15.0cm \epsfbox{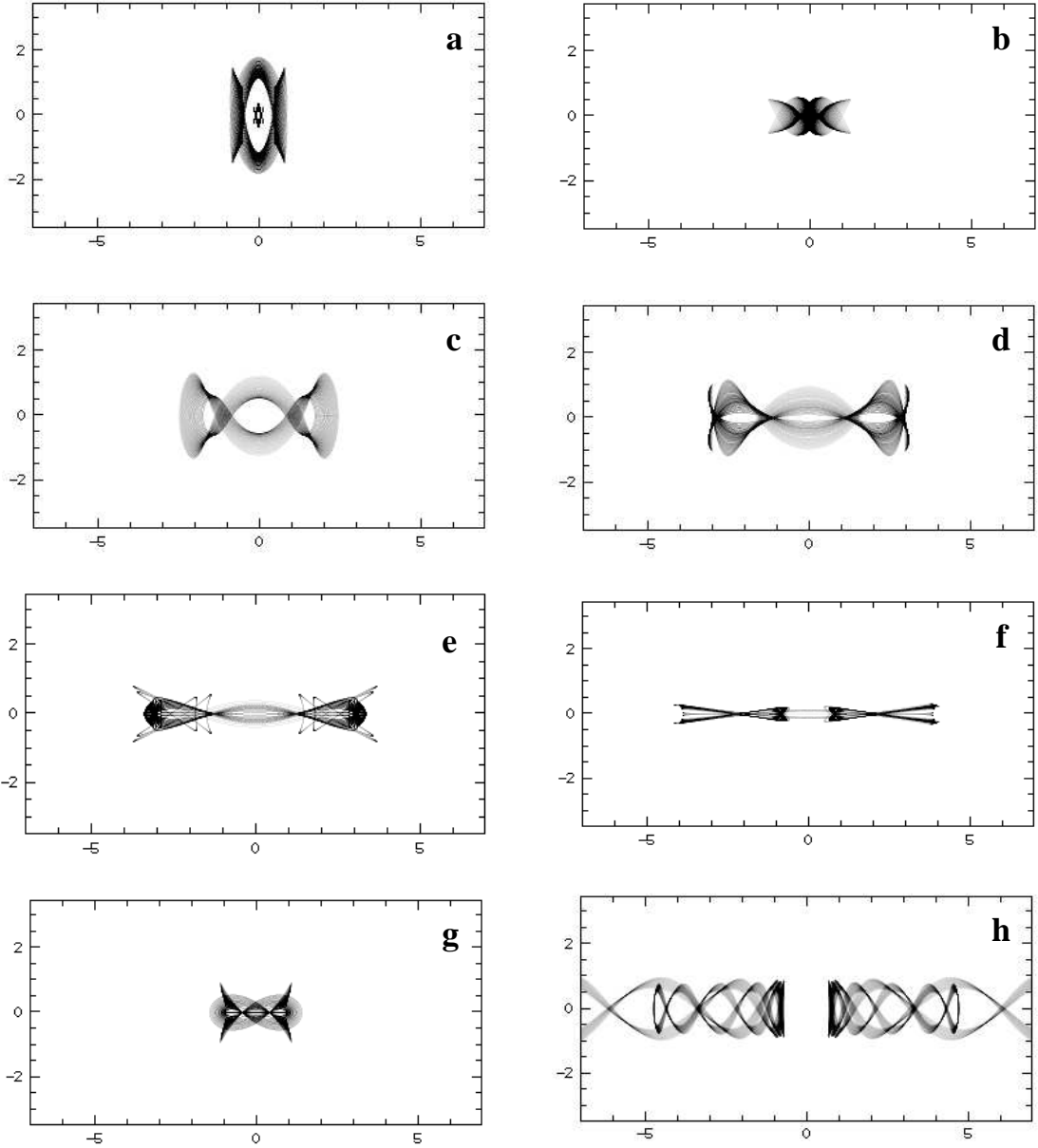}
\caption[]{
The profiles of the 3D families of model A1 viewed end-on. The layout is
as in Fig.~\ref{A1yzall}.}
\label{A1xzall}
\end{figure*}

In Fig.~\ref{A1xzall} we give the end-on, $(x,z)$, profiles for the 3D
families in model A1, i.e. we change our viewing angle by
$90\degr$ and view along the bar major axis.
Comparing Figs.~\ref{A1yzall} and \ref{A1xzall} we note that in several cases
the side-on and the end-on views of the same family have different central
morphology. The families x1v3 and x1v4 are typical examples, since the $(x,z)$
morphology of one is similar to the $(y,z)$ morphology of the other. Since in
the side-on views their profiles belong to different classes, the same holds
for their end-on views, but now x1v3 has a `\cx' and x1v4 a `\emp' central
morphology, i.e.  opposite to their side-on projections. The same happens to
the profiles of x1v7 (cf. Fig.~\ref{A1yzall}e with Fig.~\ref{A1xzall}e),
x2mul2 and the banana-like 3D orbits of family ban3v1. Thus x1v3 is the only
family with an `\cx' type end-on morphology. Viewing a barred galaxy end-on, it
does not make sense to compare the extent of the $(x,z)$ projection of a
family with the length of the bar, since the feature we study is {\em across}
the bar. So we will use the ratio of the corotation radius over the $(x,z)$
orbital length ($R_{cor}/O_{Lx}$) of the family in order to assess its 
relative length and
its extent on the galactic disc. `$R_{cor}$' indicates the corotation radius,
which in model A1 is 6.13, and `$O_{Lx}$' is the orbital length derived 
from the
projection of the family on the minor axis (similarly to the
definition of `$O_{Ly}$' for the projection on the major axis).
Another general remark is that
the $(x,z)$ projections of the profiles are in most cases smaller than their
$(y,z)$ projections. This is particularly true for the families which are
introduced in the system at low energies, i.e. x1v1, x1v3 and x1v4 shown in
Fig.~\ref{A1xzall}a, b and c respectively. Especially for x1v1, if we consider
only the part of the family before the S\ar \De transition, then its end-on
profile is tiny and thus the $R_{cor}/O_{Lx}$ ratio becomes huge. The rest of
the 3D families bifurcated from x1 have comparable `orbital lengths' in both
projections. However, we have to note that families of the x1-tree that are
bifurcated closer to corotation remain confined close to the equatorial plane
and thus are less likely to characterize the morphology of the b/p structure.

In particular for the $(x,z)$ profiles of each family
we can say the following:

The x1v3 $(x,z)$ projection (Fig.~\ref{A1xzall}b) harbours now an `\cx'
feature embedded in a boxy feature with $R_{cor}/O_{Lx} \approx 4.7$. All other
profiles are of an `\emp'-type morphology.  The x1v1 family
(Fig.~\ref{A1xzall}a) has $R_{cor}/O_{Lx} \approx 6.8$.  The
end-on projection of the orbits of this family reveals a bulge-like component
with pieces of an {\sf `X'} feature located closer to the center than in the
corresponding $(y,z)$ projection. Again one sees that the most useful orbits
of this family for building an {\sf `X'} are missing because they are complex
unstable. It is also evident that the end-on view of this family supports a
morphology that is more extended in the $z$ than in the $x$ direction, if the
outer part of the family is populated. Such a feature is not observed in
edge-on disc galaxies. That means that either such features are smoothed out
by dispersion, or that they are embedded in larger and rounder bulge
components, or that
the x1v1 orbits beyond the first S\ar \De transition are not populated.

The $(x,z)$ projection of x1v4 (Fig.~\ref{A1xzall}c) has three local maxima,
and has $R_{cor}/O_{Lx} \approx 2.5$, i.e. it is quite extended along the
equatorial plane.  This reflects the fact that many stable orbits of x1v4 in
their $(x,y)$ projections are close to hexagons with roughly equal extensions
along the $x$ and $y$ directions (see paper I). The same holds also for the
family x1v7 (Fig.~\ref{A1xzall}e), which has $R_{cor}/O_{Lx} \approx 1.8$ and
whose orbits are quite rectangular in their face-on projection. The $(x,z)$
projection of x1v7 has a third local maximum, which, however, is very close to
the equatorial plane. In this projection the x2mul2 family supports a small
peanut with $R_{cor}/O_{Lx} \approx 4.4$. The profile of family x1v9
(Fig.~\ref{A1xzall}f) is very thin in this projection as well. Finally the
3D banana-like orbits of family ban3v1 (Fig.~\ref{A1xzall}h) can contribute to
the boxiness of a central structure, by enhancing the sides of a box
at radii less
than 1~kpc, even if the central area remains empty.

To some degree, the presence of the third central local maximum seen
in the profiles in Fig.~\ref{A1yzall} and \ref{A1xzall}, depend on the
viewing angle, considered always on the equatorial plane. In the above
we considered only the side-on and end-on views, but intermediate
viewing angles can be considered in the same way.  In most cases we
examined there is a range of projection angles where the overlapping
of orbits of a family can be responsible for the morphology of a
continuous `\cxs' feature inside a box. In other words, by changing
the viewing angle we can minimize the importance of the third local
maximum in $z$, which is not consistent with a peanut morphology, and
at the same time bring the two separated pieces of an {\sf `X'}
closer. However, one should not overestimate the role of rotation,
because the range of viewing angles that reproduce the {\sf `X'}
morphology is narrow. In the case of the x1v1 family of model A1, a
{\sf `X'} feature due to rotation is discernible only in an angle range
$\Delta \theta \lid 15\degr$ close to the $(x,z)$-projection.

In Table~\ref{tab:A1tab} we summarize the properties of the edge-on orbital
profiles in model A1. The symbols used for describing the central morphologies
are explained in the text, except for `][' used for the central end-on
morphology of the 3D banana-like orbits of the family ban3v1. The two branches
of this family appear at the sides of the bar, symmetric with respect to its
major axis, and leave a totally empty region
at the center of the end-on profile.
\begin{table*}
\caption[]{Properties of edge-on profiles in model A1. At the top row we
summarize the properties of the model. G is the gravitational constant,
M$_D$, M$_B$, M$_S$ are the masses of the disk, the bar and the
bulge respectively, $\epsilon_s$ is the scale length of the bulge, \O$_{b}$ is
the pattern speed of the bar, $E_j$(r-IILR) is the Jacobian for the
inner radial ILR, $E_j$(v-ILR) is the Jacobian for the vertical 2:1
resonance, $R_c$ is the corotation radius, and $B_L$ is the longest projection
of bar supporting orbits on the semi-major axis. The comment characterizes
briefly the model. The successive columns of the main table give the name of the
family, the vector of the initial conditions on the Poincar\'{e}
surface of section defined by $y_0=0$ and $\dot{y_0}>0$, essentially
describing if we have 
a vertical bifurcation of x1 in $z$ or $\dot{z}$, the $B_L/O_{Ly}$ ratio, a
symbol indicating the central morphology supported by the orbit  in the side-on
view (s/o), the $R_{cor}/O_{Lx}$ ratio, a symbol indicating the central
morphology supported by the orbit in the end-on view (e/o), and finally a
comment about special features of the particular profile. }
\label{tab:A1tab}
\begin{center}
\begin{tabular}{ccccccccccc}
model name& GM$_D$ & GM$_B$ & GM$_S$ & $\epsilon_s$ & \O$_{b}$ & $E_j$(r-IILR)
&$E_j$(v-ILR)
& $R_c$ & $B_L$ &comments\\
\hline
A1 & 0.82 &  0.1  & 0.08 & 0.4 &  0.0540 & -0.441 & -0.360&    6.13 & 4.5 &
fiducial \\
\hline
\end{tabular}
\begin{tabular}{ccccccc}
family & initial conditions & $B_L/O_{Ly}$ & s/o shape & $R_{cor}/O_{Lx}$ & e/o
shape  & comments\\
\hline
x1v1   & $(x,z,0,0)$ &  4.1/2.1    & \emp & 41/6.8&  \emp & two stable
parts (see text)\\
x1v3   & $(x,z,0,0)$        &  1.2 & \emp & 4.7 &  \cx  &  \\
x1v4   & $(x,0,0,\dot{z})$  &  1.3 & \cx  & 2.5 &  \emp &  \\
x1v5   & $(x,z,0,0)$        &  1.2 & \emp & 2.0 &  \emp &  \\
x1v7   & $(x,0,0,\dot{z})$  &  1.2/0.9 & \cx  & 1.8/1.7 & \emp & partly not
bar-supporting \\
x1v9   & $(x,z,0,0)$  &  --  & \emp & 1.5 &  \emp & not bar-supporting\\
x2mul2 & $(x,z,0,0)$        &  9   & \cx  & 4.4 &  \emp & along the minor axis
of the bar\\
ban3v1 & $(x,0,0,\dot{z})$  &  --  & \cx  & --  &  ][   & not bar-supporting\\
\hline
\end{tabular}
\end{center}
\end{table*}

\subsection{Model A2}
Model A2 differs from A1 only in the pattern speed (paper II), which
is so much slower that its inner Lindblad resonance is
roughly at the radius where model A1 has its corotation.

\begin{figure*}
\epsfxsize=14.0cm \epsfbox{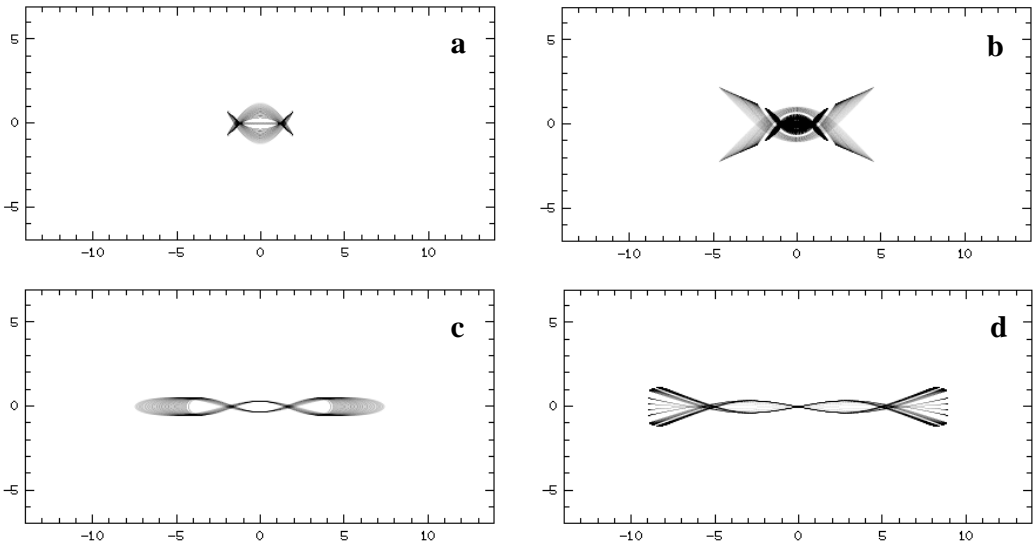}
\caption[]{The side-on profiles of the main relevant families of model
  A2. Each panel corresponds to 
another family: (a) x2v1, (b) x1v1, (c) x1v3 and (d) x1v4. In model A2
corotation
  is at 13.24.}
\label{A2yzall}
\end{figure*}

In Fig.~\ref{A2yzall} we give
the $(y,z)$ projections of the four families that can essentially affect
the appearance of the vertical profile of the model.
It is evident that in this projection b/p bulges can be associated with either
of the families x2v1 (Fig.~\ref{A2yzall}a) and x1v1 (Fig.~\ref{A2yzall}b). The
other two families have profiles that remain close to the equatorial plane
while reaching in this plane close to corotation and thus can be associated only
with the vertical structure in the outer parts of the bar. This is even more
the case for the rest of the stable 3D families of model A2 described in paper
II (x1$^{\prime}$v4, x1$^{\prime}$v5), as they remain very close to the
equatorial plane and do not influence the edge-on morphology of the model.
The projection of the bar-supporting orbits in model A2 on the semi-major axis
reaches a distance from the center $r\approx 9.4$. So we have $B_L/O_{Ly}
\approx 2.1$ for the x1v1 family and $B_L/O_{Ly} \approx 4.7$ for x2v1
(the x2v1 orbits are elongated along the minor axis of the main bar).  The
superposition of x1v1 orbits will give to the system a very sharp and
continuous {\sf `X'} feature despite the fact that individual orbits do not
support this morphology, even if we take into account both branches which are
symmetric with respect to the equatorial plane. The reason for this is that 
the gap due to the complex unstable part at the S\ar \D \ar S transition 
is negligible. Family x2v1 has a `\emp' feature in the center.

\begin{figure*}
\epsfxsize=14.0cm \epsfbox{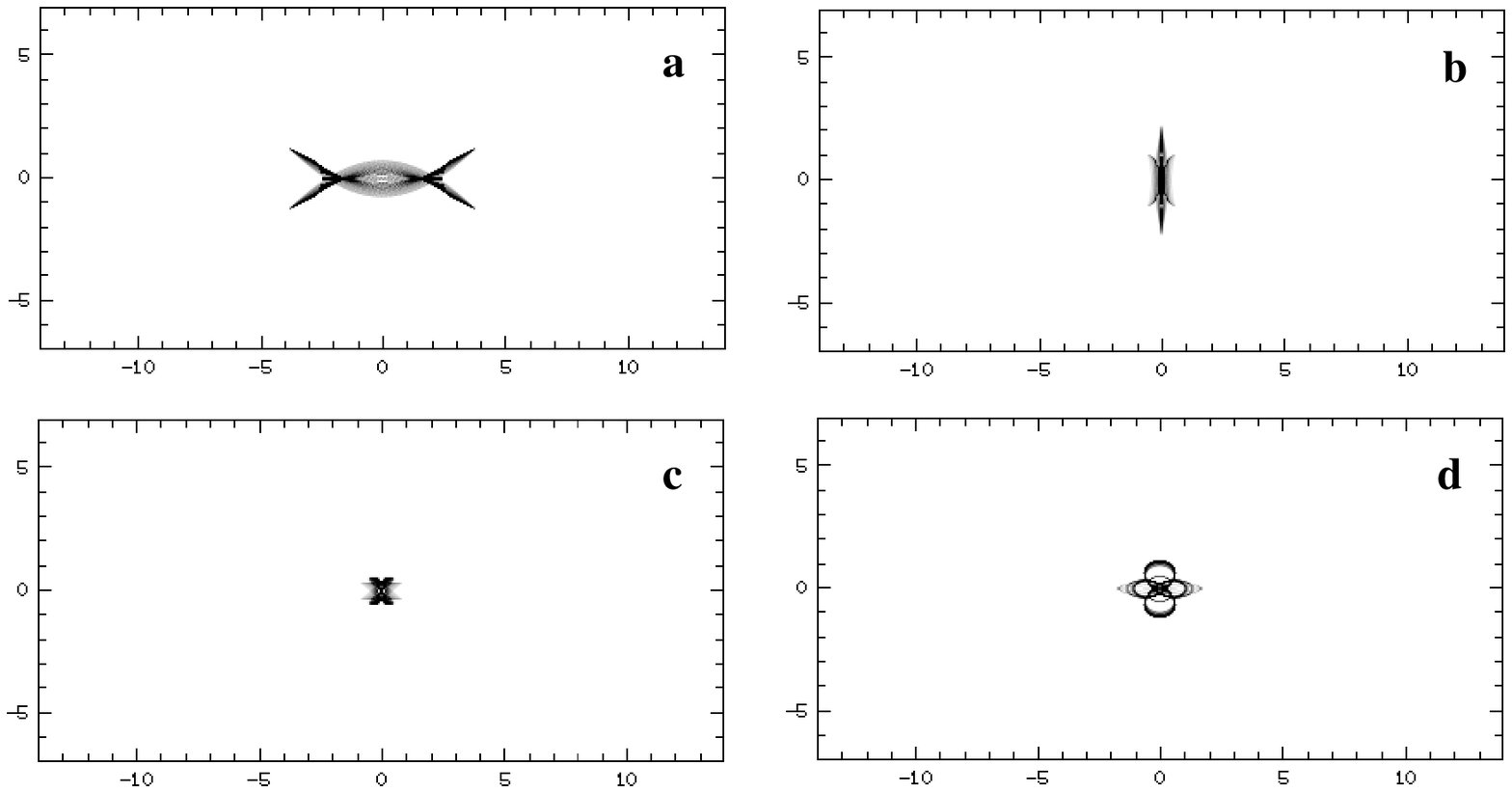}
\caption[]{The end-on profiles of the main relevant families of model A2.
The layout is as in Fig.~\ref{A2yzall}.}
\label{A2xzall}
\end{figure*}

If we rotate the model by $\pi/2$ the families will give the projected
profiles depicted in Fig.~\ref{A2xzall}.
In this projection also it has a `\emp'
central part (Fig.~\ref{A2xzall}a) and remains relatively close to the
equatorial plane. The x1v1 profile in this projection almost vanishes
squeezed on the rotational axis (Fig.~\ref{A2xzall}b). The x1v3
profile can be described as a tiny {\sf `x'}, while that of x1v4 can be 
described as a
quadrupole feature made of the overlapping of an `8' and an `$\infty$'
symbol. We remind here that in all these profiles we consider both,
symmetric with respect to the equatorial plane, branches of each
family. If due to
the presence of a an external factor (e.g. a companion) only one
branch is populated then the above symmetry will break, and instead
of e.g a tiny  {\sf `x'} feature, in the case of x1v3, this family will
support the presence of a tiny `$\sim$' feature.

The properties of the profile of each family are summarized in
Table~\ref{tab:A2tab}. Topologically, the central morphology of a family is of
course similar in all models. However, in the tables we prefer to refer to
symbols that characterize the specific morphology in each case. Therefore we
describe in model A2 (Table~\ref{tab:A2tab}) the central side-on x1v1
morphology with the symbol {\sf `X'}, while to the end-on central morphologies
of the families x1v1, x1v3 and x1v4 we attribute the symbols `$|$', {\sf `x'}
and `8/$\infty$', respectively.
\begin{table*}
\caption[]{Properties of edge-on profiles in model A2. The columns are as in
  Table 1.}
\label{tab:A2tab}
\begin{center}
\begin{tabular}{ccccccccccc}
model name& GM$_D$ & GM$_B$ & GM$_S$ & $\epsilon_s$ & \O$_{b}$ & $E_j$(r-IILR)
&$E_j$(v-ILR)
& $R_c$ & $B_L$ & comments\\
\hline
A2 & 0.82 &  0.1  & 0.08 & 0.4 &  0.0200 & -0.470 & -0.357&   13.24 & 9.4 &
slow bar \\
\hline
\end{tabular}
\begin{tabular}{ccccccc}
family & initial conditions & $B_L/O_{Ly}$ & s/o shape & $R_{cor}/O_{Lx}$ & e/o
shape  & comments\\
\hline
x2v1   & $(x,z,0,0)$        &  4.7  & \emp & 3.6 &  \emp & along the minor
axis of the bar\\
x1v1   & $(x,z,0,0)$        &  2.1  & {\sf X} & 26.5 &  $|$ & \\
x1v3   & $(x,z,0,0)$        &  1.3 & \emp & 16.6 & {\sf x } & \\
x1v4   & $(x,0,0,\dot{z})$  &  1.1 & \emp & 7.8 &  8\&$\infty$ &  \\
\hline
\end{tabular}
\end{center}
\end{table*}

\subsection{Model A3}
\begin{figure}
\epsfxsize=9.0cm \epsfbox{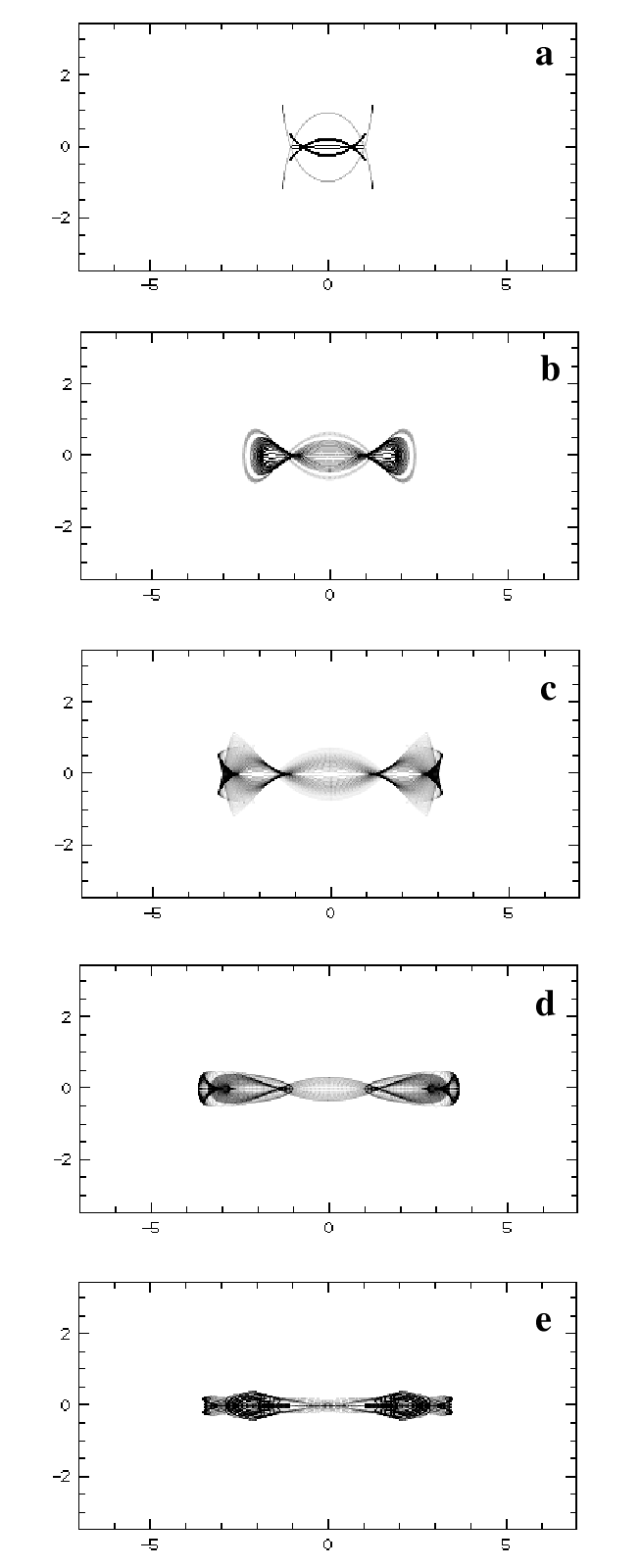}
\caption[]{The side-on profiles of the  families of model A3.
Each panel corresponds to
another family: (a) x1v1, (b) x1v3, (c) x1v5, (d) x1v8 and (e)
  the 3D bifurcation of q0, i.e. q0v1.  In model A3 corotation is at 4.19. }
\label{A3yzall}
\end{figure}

Model A3 is the same as model A1, except that its bar
is rotating faster (paper II).
The main 3D families which could build its
vertical structure are x1v1, x1v3, x1v5, x1v8 and q0v1, the 3D bifurcation of
family q0 (paper II). Their $(y,z)$ profiles are given in
Fig.~\ref{A3yzall}.

At the side-on projections only q0v1 has at
the center a `\cx' morphology. However, this is again a 4:1 radial
resonance family whose orbits have four big loops and thus does not
directly support the bar. Also its range of existence is narrow. All
other profiles have a central region that can be described with the
symbol `\emp'.

As we see in Fig.~\ref{A3yzall}a the x1v1 profile is characterized by
a large empty part because of the complex unstable region. In the
stable part beyond the \De \ar S transition the family gives
proportionally only a few orbits in relatively low heights above the
equatorial plane. For energies somewhat larger than the energy of the
outermost x1v1 orbit plotted in Fig.~\ref{A3yzall}a the orbits reach
vertical distances with $|z| = 2$, turning simultaneously towards
lower $|y|$ values; they are not given in Fig.~\ref{A3yzall}a.  The
rest of the side-on profiles do not have essential differences from
what we encountered in the fiducial model.

\begin{figure}
\epsfxsize=8.7cm \epsfbox{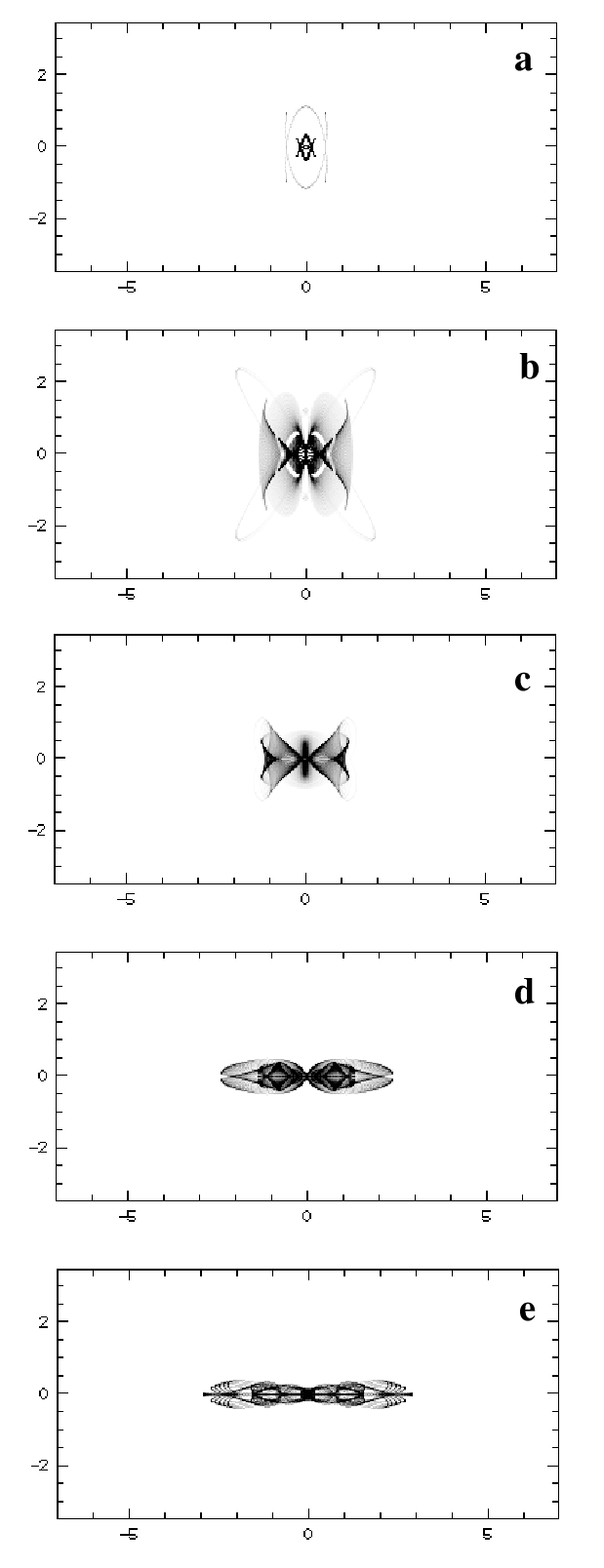}
\caption[]{The end-on profiles of the families of model A3.
The layout is as in Fig.~\ref{A3yzall}. }
\label{A3xzall}
\end{figure}

Contrary to the side-on, the end-on projections (Fig.~\ref{A3xzall}) of the
families x1v3 (Fig.~\ref{A3xzall}b)
and x1v5 (Fig.~\ref{A3xzall}c) support the presence of b/p features. The
x1v5 profile is characterized in addition by two local enhancements of 
the surface density along the minor axis which are symmetric with 
respect to the center. The  size of these b/p structures is small compared to the
corotation radius and gives $R_{cor}/O_{Lx}=3.5$ both for the x1v3, and
for the x1v5 family.
x1v8 (Fig.~\ref{A3xzall}d) and especially
q0v1 (Fig.~\ref{A3xzall}e) are less important for the end-on vertical
structure of the model. All of the families, except for x1v1 which
anyway vanishes in the end-on projection (Fig.~\ref{A3xzall}a), give
profiles with a `\cx' central morphology. In conclusion, model A3 can
give only confined boxy profiles when viewed end-on.

The properties of the profile of each family are summarized in
Table~\ref{tab:A3tab}. The two ratios given in
Table~\ref{tab:A3tab} in the $B_L/O_{Ly}$, as well as in the $R_{cor}/O_{Lx}$,
columns,
refer to the x1v1 orbits as plotted in Figs.~\ref{A3yzall}a and
\ref{A3xzall}a. By adding a question mark in a parenthesis after the
symbol `{\sf X}', which characterizes the central end-on morphology of the
family x1v5, we want to underline that, although the family supports the `{\sf
  X}' morphology, the confined extent of its end-on projection will apparently
lead just to a boxy profile.
\begin{table*}
\caption[]{Properties of edge-on profiles in model A3. The columns are as in Table 1.}
\label{tab:A3tab}
\begin{center}
\begin{tabular}{ccccccccccc}
model name& GM$_D$ & GM$_B$ & GM$_S$ & $\epsilon_s$ & \O$_{b}$ & $E_j$(r-IILR)
&$E_j$(v-ILR)
& $R_c$ & $B_L$ & comments\\
\hline
A3 & 0.82 &  0.1  & 0.08 & 0.4 &  0.0837 & -0.390 & -0.364&    4.19 & 3.8 &
fast bar \\
\hline
\end{tabular}
\begin{tabular}{ccccccc}
family & initial conditions & $B_L/O_{Ly}$ & s/o shape & $R_{cor}/O_{Lx}$ & e/o
shape  & comments\\
\hline
x1v1   & $(x,z,0,0)$        &  3.8/3.2  & \emp & 16.8/7.0 &  \emp & \\
x1v3   & $(x,z,0,0)$        &  1.6  & \emp & 3.5 &  \cx  &  \\
x1v5   & $(x,z,0,0)$        &  1.2 & \emp & 3.5 & {\sf X}(?) & \\
x1v8   & $(x,z,0,0)$        &  1.0 & \emp & 1.7 &  \cx & \\
q0v1   & $(x,z,\dot{x},\dot{z})$ &  1.1 & \cx  & 1.4 &  \cx & not directly
bifurcated from x1\\
\hline
\end{tabular}
\end{center}
\end{table*}

\subsection{Model B}
\begin{figure*}
\rotate[r]{
\epsfxsize=3.1cm \epsfbox{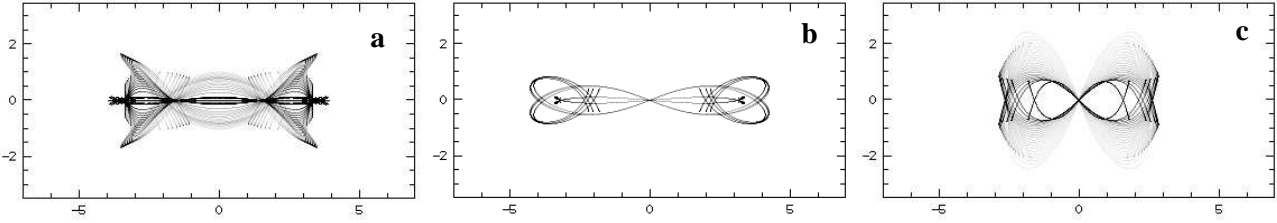}}
\caption[]{The side-on profiles of the  families of model B.
Each panel corresponds to
another family: (a) x1v5 and x1v5$^{\prime}$, (b) x1v7, (c) z3.1s. Model B has
  neither radial nor vertical 2:1 resonances and its corotation is at
  6.}
\label{Myz}
\end{figure*}

\begin{figure*}
\rotate[r]{
\epsfxsize=3.5cm \epsfbox{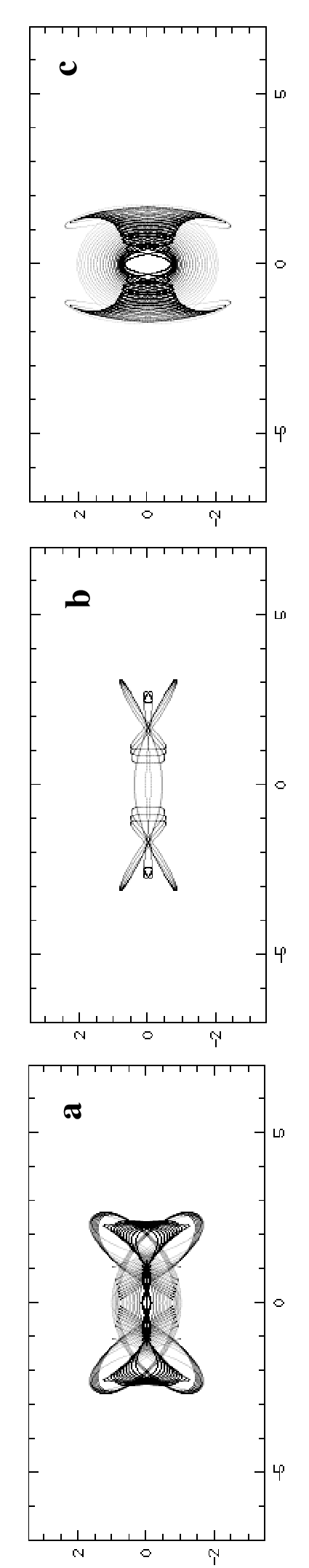}}
\caption[]{The end-on profiles of the families of model B.
The layout is as in Fig.~\ref{Myz}.}
\label{Mxz}
\end{figure*}

As described in paper II, model B has the same total mass as the
fiducial one but
lacks an explicit bulge component. It
has neither radial, nor vertical 2:1 resonances, and its bar rotates as fast
as that of model A1. As there are no x1v1 orbits close to the center to
populate a boxy structure, and there is no Plummer sphere to
  influence  the central
dynamics of the model, one could have expected to have more
flat edge-on profiles. This, however, is not the case, as we see in
Fig.~\ref{Myz} for the
profiles along the major axis and Fig.~\ref{Mxz} for the profiles
along
the minor axis of
the bar.

The families that could be related with boxy vertical structures in this model
are the two families related to the vertical 4:1 resonance (x1v5,
x1v5$^{\prime}$), x1v7, and z3.1s. As we have seen in paper II, the vertical
bifurcations of x1 that give 3D orbits are x1v5, x1v5$^{\prime}$, and x1v7;
x1v5 being the first vertical x1 bifurcation in this model. Their side-on
profiles give the boxy structures, which are depicted in Fig.~\ref{Myz}a for
x1v5 and x1v5$^{\prime}$ together ($B_L/O_{Ly} \approx 1.1$), and
Fig.~\ref{Myz}b for x1v7 ($B_L/O_{Ly} \approx 1)$. The outer orbits of x1v7,
however, as in the fiducial case, are characterized in their face-on
projection by an extreme `butterfly' morphology {\em surrounding} all other
orbits that support the bar.  Furthermore, the inner orbits remain very close
to the equatorial plane. Thus the significance of this and of other similar 
families
is small. x1v5$^{\prime}$ orbits also have at their largest energies a
`butterfly' morphology as well. Their $(x,y)$ projections, however, are
surrounded by the usual rectangular orbits at the 4:1 region (paper IV). Also
we note that, although the combined side-on profile of the two vertical 4:1
resonance families has three local maxima above the equatorial plane, as does
x1v5 in model A1 (Fig.~\ref{A1yzall}d), the central local maximum in model B
is clearly lower than those on the sides, and this leads to a b/p morphology
similar to that observed in edge-on galaxies with b/p profiles. Most
interesting in this model is the side-on profile of the z3.1s family
(Fig.~\ref{Myz}c). As we have seen in paper II, this family is not part of the
x1 forest. It is connected to the z-axis orbits described three times.
Fig.~\ref{Myz}c shows that its profile supports a boxy structure, as well as
the presence of a central `\cx' morphology and thus makes a perfect peanut.
The peanut has $B_L/O_{Ly} \approx 1.4$, while if we consider only the orbits
that make the strong part of the `\cx' we have $B_L/O_{Ly} \approx 2$.

The profiles along the minor axis can also be boxy (Fig.~\ref{Mxz}). x1v5 and
x1v5$^{\prime}$ build a well-defined peanut along the {\em minor} axis with
$R_{cor}/O_{Lx} \approx 2.3$. The z3.1s profile could give either a boxy or a
rounder bulge-like feature, depending on the degree of participation of the
orbits with the largest energies, which give the horn-like morphology of this
profile. If we include the high energy orbits, the profile has the form of a
roundish bulge-like feature, slightly elongated along the rotational axis.

The end-on profile of family x1v7 is confined close to the equatorial plane
(Fig.~\ref{Mxz}b), with a $R_{cor}/O_{Lx}$ ratio about 1.9. Thus, in the
presence of
the orbits of the two other families it does not characterize the morphology
of the central area of the model even in this projection.

The properties of the profile of each family of model B are
summarized in Table~\ref{tab:Btab}.
\begin{table*}
\caption[]{Properties of edge-on profiles in model B. The columns are as in
  Table 1.}
\label{tab:Btab}
\begin{center}
\begin{tabular}{ccccccccccc}
model name& GM$_D$ & GM$_B$ & GM$_S$ & $\epsilon_s$ & \O$_{b}$ & $E_j$(r-IILR)
&$E_j$(v-ILR)
& $R_c$ & $B_L$ & comments\\
\hline
B  & 0.90 &  0.1  & 0.00 & --  &  0.0540 &  --    &  --   &    6.00 & 3.9 &
no bulge \\
\hline
\end{tabular}
\begin{tabular}{ccccccc}
family & initial conditions & $B_L/O_{Ly}$ & s/o shape & $R_{cor}/O_{Lx}$ & e/o
shape  & comments\\
\hline
x1v5/x1v5$^{\prime}$    & $(x,z,0,0)$        &  1.1     & \emp & 2.3 &  \emp &
\\
x1v7   & $(x,0,0,\dot{z})$  &  1       & \cx  & 1.9 &  \emp &  \\
z3.1s  & $(x,0,0,\dot{z})$  &  1.4/2.0 & \cx  & 3.5 & \emp & not part of the
x1-tree\\
\hline
\end{tabular}
\end{center}
\end{table*}

\subsection{Model C}
Model C has a vertical, but no radial, 2:1 resonance.  The
edge-on and end-on profiles in this model are in general similar to
those encountered in model A1. There is, however, a notable exception
and this refers to the family x1v1, bifurcated at the vertical 2:1
resonance. x1v1, in this particular model, has an S\ar \De transition
at large energies (paper II). It thus remains practically always stable and
its orbits which do not reach large distances over the equatorial
plane, could populate a boxy feature. Comparison of model C with model A1
demonstrates the role of the complex instability in the b/p profiles made
by x1v1 orbits. In the present case the {\sf `X'} feature is not
interrupted. This difference will become apparent in Section~5, where we
compare the various morphologies of the models. The properties of the
model and of the x1v1 family are summarized in Table~\ref{tab:Ctab}.
\begin{table*}
\caption[]{Properties of edge-on profiles in model C. The columns are as in Table 1.}
\label{tab:Ctab}
\begin{center}
\begin{tabular}{ccccccccccc}
model name& GM$_D$ & GM$_B$ & GM$_S$ & $\epsilon_s$ & \O$_{b}$ & $E_j$(r-IILR)
&$E_j$(v-ILR)
& $R_c$ & $B_L$ & comments\\
\hline
C  & 0.82 &  0.1  & 0.08 & 1.0 &  0.0540 &  --    & -0.364&    6.12 & 4.2 &
extended bulge \\
\hline
\end{tabular}
\begin{tabular}{ccccccc}
family & initial conditions & $B_L/O_{Ly}$ & s/o shape & $R_{cor}/O_{Lx}$ & e/o
shape  & comments\\
\hline
x1v1   & $(x,z,0,0)$ &  2.3  & {\sf X} & 6.8 &  \emp &  \\
\hline
\end{tabular}
\end{center}
\end{table*}

\subsection{Model D}
\begin{figure}
\epsfxsize=8cm \epsfbox{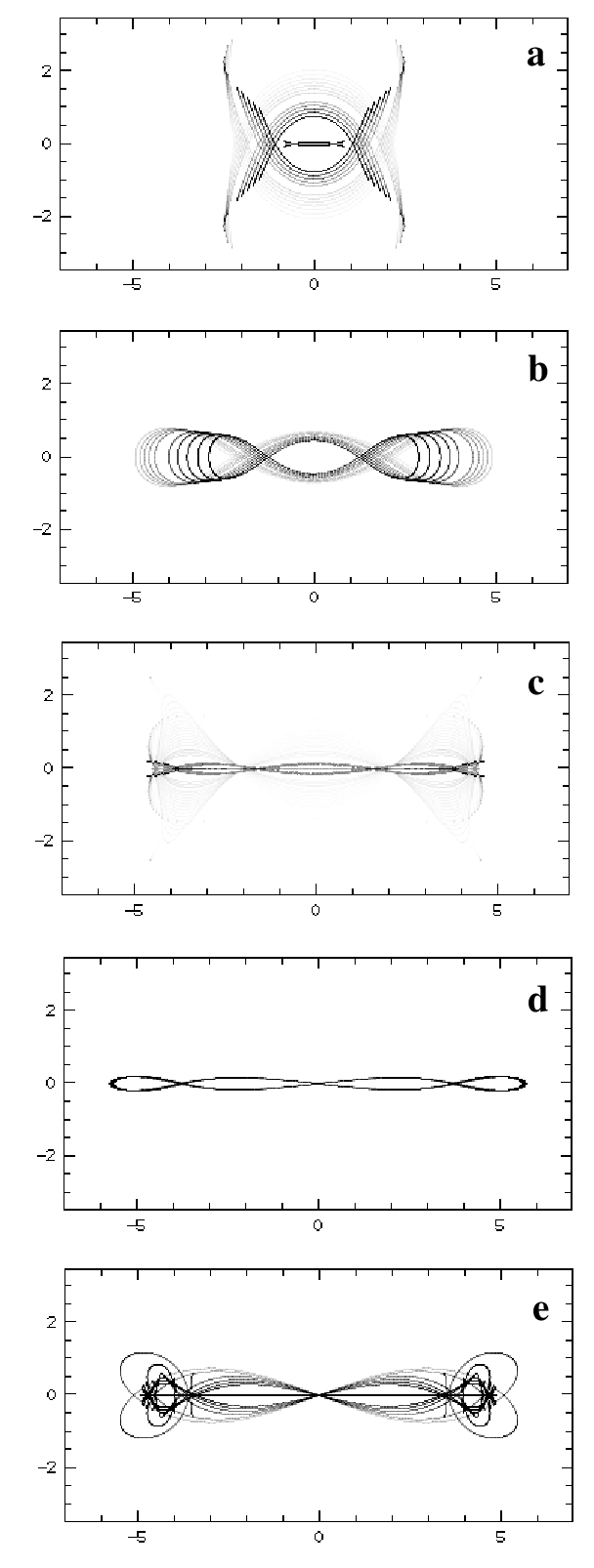}
\caption[]{The side-on profiles of the  families of model D.
Each panel corresponds to
another family: (a) x1v1,
(b) x1v3, (c) x1v5 and x1v5$^\prime$, (d) x1v6, (e) x1v7. In model D
corotation is at 6.31.}
\label{Hyz}
\end{figure}

\begin{figure}
\epsfxsize=8cm \epsfbox{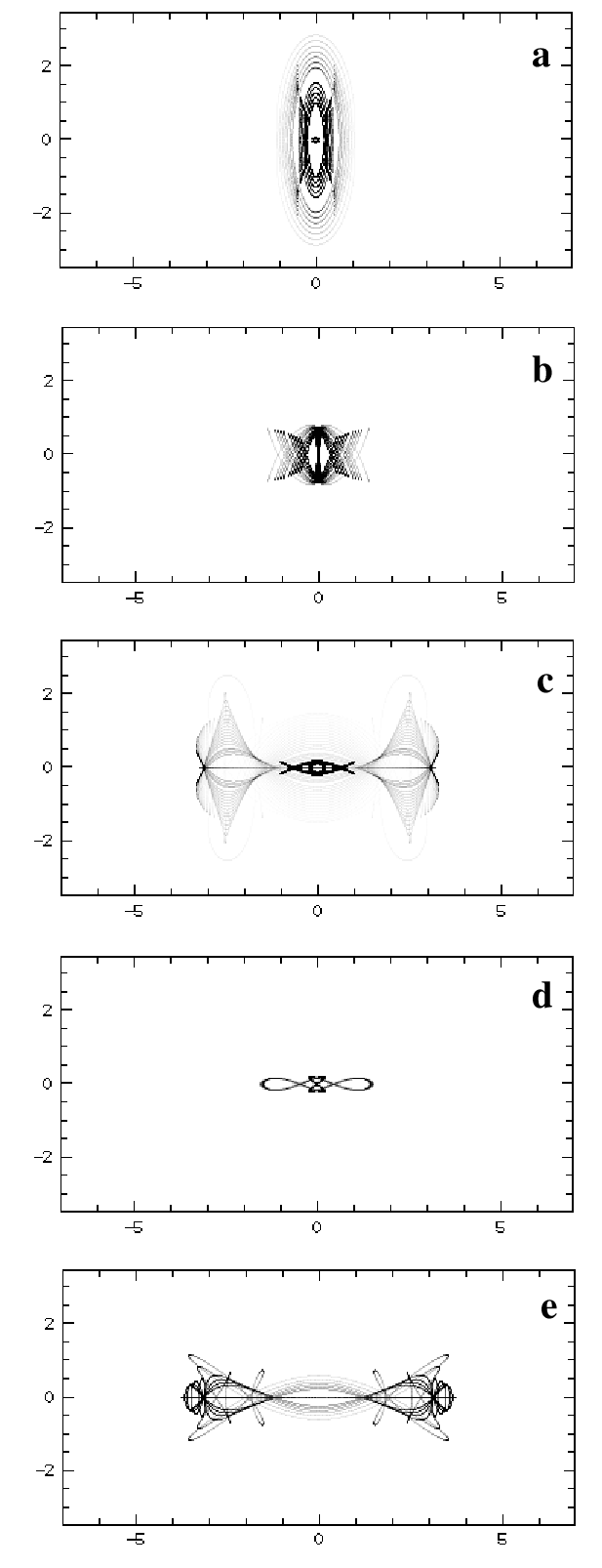}
\caption[]{The end-on profiles of the families of model D.
The layout is as in Fig.~\ref{Hyz}.}
\label{Hxz}
\end{figure}

A strong bar case is described in model D. The mass of the bar in model A1 is
doubled at the expense of the disc mass, so that the total mass of the system
is kept constant. Again we have collected the side-on and end-on profiles 
of the
families in two figures; Fig.~\ref{Hyz} for the $(y,z)$ and Fig.~\ref{Hxz} for
the $(x,z)$ profile. In this case we have, as in
model B, two families associated with the vertical 4:1 resonance (paper II).

In the side-on views families x1v1, x1v3, x1v5 and x1v5$^{\prime}$ have
central regions characterized by an `\emp' feature.  In Fig.~\ref{Hyz}a we
observe that the main part of the profile consists of x1v1 orbits which are
nested one inside the other, in such a way as to build outside of the central
empty region parts of the four sides of an {\sf `X'} feature. The difference
with the corresponding profile of the fiducial case (Fig.~\ref{A1yzall}a) is
that the branches of the {\sf `X'} are now more extended. In
Fig.~\ref{A1yzall}a the pieces of the {\sf `X'} branches emerging out of the
empty region reach $|z| \approx 1.5$ in a distance of $\Delta y \approx 0.5$,
while in Fig.~\ref{Hyz}a in a distance of $\Delta y \approx 1$. In model D,
the family x1v1 undergoes a S\ar \D \ar S transition, like in model A1. Model D,
however, is more suitable than model A1 to describe b/p features because in
this model
the orbits needed for building the peanut are stable and not complex unstable.
As the bar is more massive, the potential well is deeper than that of model
A1.  Thus the family x1v1 is bifurcated at a smaller energy than in model A1,
becomes complex unstable, and when it turns again stable the orbits have
relatively low $\overline{|z|}$.  The ratio $B_L/O_{Ly}$ is $\approx 2.4$. A
secondary gap in the succession of the x1v1 orbits that can be seen in
Fig.~\ref{Hyz}a, is due to a small instability strip (Fig.16 in paper II) at
which x1v1 has a simple unstable part. This gap, however, is bridged by a
stable bifurcation of x1v1, called
x1v1.1. The x1v3 orbits (Fig.~\ref{Hyz}b) reach lower $|z|$ values than those
of x1v1. They could affect the vertical profile, but not in the central
region. Finally, we have the two families bifurcated at the vertical 4:1
resonance area:  x1v5 with elliptical-like orbits with loops along the major
axis
at their face-on projections and x1v5$^{\prime}$ with rectangular like orbits
in the $(x,y)$ projection. In the side-on profile the x1v5 orbits remain close
to the equatorial plane, while the orbits of x1v5$^{\prime}$ reach large
${|z|}$. In both cases the contribution of the family is not important for the
vertical structure of this model because of the mean deviation of the orbits
from the equatorial plane. The families x1v6 and x1v7 have in the central
regions of their side-on profiles a `\cx' feature. In this model x1v6 has
some stable representatives, but they remain always close to the $z=0$ plane
and its $B_L/O_{Ly}$ ratio is close to 1.
x1v7 gives a b/p vertical structure with a ratio $B_L/O_{Ly}
\approx 1.2$. This means that most of the bar participates in the b/p
morphology.

Viewed end-on, the x1v1 profile gives a bulge-like feature, rather
elongated along the $z$ axis of rotation. x1v7, although it shows in this
projection  the three $z$ maxima morphology, can be described as
boxy or peanut-shaped, because the two clumps at the sides are much
denser than the middle one. Finally, in this projection, family x1v3
could provide a confined boxy feature.

The properties of
the model and of the profiles are summarized in Table~\ref{tab:Dtab}.
\begin{table*}
\caption[]{Properties of edge-on profiles in model D. The columns are as in
  Table 1.}
\label{tab:Dtab}
\begin{center}
\begin{tabular}{ccccccccccc}
model name& GM$_D$ & GM$_B$ & GM$_S$ & $\epsilon_s$ & \O$_{b}$ & $E_j$(r-IILR)
&$E_j$(v-ILR)
& $R_c$ & $B_L$ & comments\\
\hline
D  & 0.72 &  0.2  & 0.08 & 0.4 &  0.0540 & -0.467 & -0.440&    6.31 & 5.7 &
strong bar \\
\hline
\end{tabular}
\begin{tabular}{ccccccc}
family & initial conditions & $B_L/O_{Ly}$ & s/o shape & $R_{cor}/O_{Lx}$ & e/o
shape  & comments\\
\hline
x1v1   & $(x,z,0,0)$        &  2.4  & \emp /{\sf X} & 6.3 &  \emp &  \\
x1v3   & $(x,z,0,0)$        &  1.2  & \emp & 4.9 & {\sf x} & \\
x1v5/x1v5$^{\prime}$    & $(x,z,0,0)$        &  1.2 & \emp & 1.9 & \emp  & \\
x1v6   & $(x,0,0,\dot{z})$  &  1.0 & \cx & 4.2 &  \emp & \\
x1v7   & $(x,0,0,\dot{z})$  &  1.2 & \cx & 1.7 &  \emp & \\
\hline
\end{tabular}
\end{center}
\end{table*}
In effect, in the box of the side-on morphology of x1v1, which is
essentially of `\emp' type, we can see segments of a `{\sf X}'
feature. Segments of `{\sf X}' can also be seen in the boxy end-on profile of
the family x1v3. However due to its large $R_{cor}/O_{Lx}$ ratio, the
addition of
dispersion in the velocities of the orbits will result in a boxy
feature. These peculiarities are indicated by the symbols we use in
Table~\ref{tab:Dtab}.

\section{Composite profiles}
So far we presented separately the building blocks that each
family provides to the vertical structure of a model. However, the
vertical structure of a real galaxy could be specified by more than
one family. So we created composite profiles as well, taking into
account all, or some, of the families of the model in each case. In
this kind of representations the images of the profiles we have seen
until now are added together to give a single image.

\begin{figure}
\epsfxsize=8.0cm \epsfbox{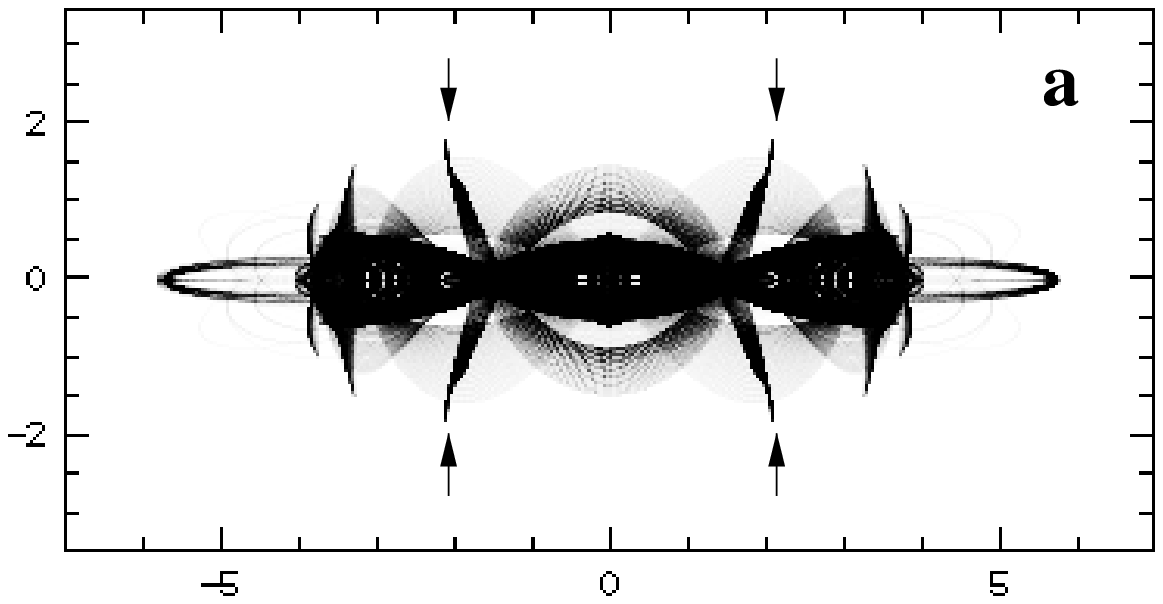}
\epsfxsize=8.0cm \epsfbox{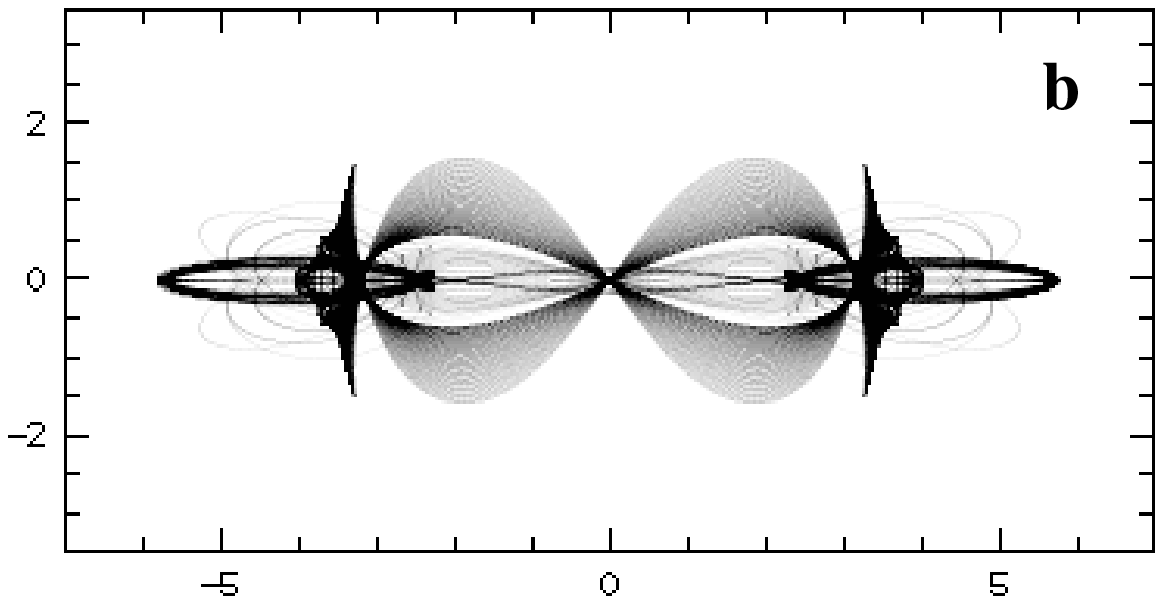}
\caption[]{Composite side-on, $(y,z)$, vertical profiles of model A1. (a) All
3D families except the banana-like orbits are considered. (b) The
profile consists of families x1v4, x1v7 and x1v9.}
\label{Sc1yz}
\end{figure}

In Fig.~\ref{Sc1yz}a, we consider all families of model A1 presented
in Fig.~\ref{A1yzall}, except for the banana-like orbits.
If we select just the families x1v4, x1v7 and x1v9 we
obtain the profile we see in Fig.~\ref{Sc1yz}b.
In principle there is no reason to include in a system all families (or all
orbits of a family) populated. Internal and external factors, like resonance
widths, formation history, influence of companions etc. decide which of them
will be populated in a real galaxy. E.g. in Fig.~\ref{Sc1yz}a a boxy bulge
could be built inside the `borders' indicated by arrows, which are imposed by
the x1v1 orbits. However, if for some reason x1v1 is not populated and the
bar, apart from the stable members of the 2D x1 orbits, is made out of the
families x1v4, x1v7 and x1v9, we have again a b/p profile, as we see in
Fig.~\ref{Sc1yz}b.  The main difference between the two configurations, that
would reflect in the photometry of edge-on galaxies, is the $B_L/O_{Ly}$ ratio
or a similar indicator. In the present case a boxy bulge due to the x1v1
family would give $B_L/O_{Ly} \approx 2.1$, while the profile that is dominated
by x1v4 has $B_L/O_{Ly} \approx 1.3$.

The family bifurcated from x1 at lowest energy, if populated, will
characterize qualitatively the boxiness of a model's profile, because it
determines whether we will have a peanut, a boxy, or a flat edge-on profile.
The reason for this is that the orbits of a 3D family which bifurcated
from x1 at a $n:1$ vertical resonance, have in general larger $\overline{|z|}$
than the orbits of the family bifurcated at the $(n+1):1$ vertical
resonance.
The family x1v3 can be an exception to this rule in some models (e.g. model
A1), if it undergoes a S\ar \De transition when its orbits are still confined
close to the equatorial plane without becoming stable for larger
energies. This stability behaviour, however, has been encountered only in
specific models and is not the general rule. In our models we found profiles
whose vertical extent
decreases as we move away from the center of the galaxy. They present
relatively abrupt height changes at distances from the center which are
characteristic for each family. So, the orbits of a given family reach 
a maximum distance from the
center along an axis. This is especially
discernible in the case of the x1v1 family. These `stair-type' edge-on profiles
have been encountered in 3D spiral potentials \cite{pg96} and in almost
axisymmetric disc models as well (Patsis, Athanassoula, Grosb{\o}l et al.
2002a).

Blurred smoothed images are another informative representation of the
profiles. 
They are created by applying a gaussian smoothing
filter on the images using the corresponding MIDAS command. The smoothing
radius, i.e the number of pixels 
``around'' each
central pixel, is 9 in both directions of the image. So we have a
19$\times$19 pixels smoothing neighbourhood.
The mean and sigma values that have been used are 9 and 6 pixels respectively,
again in both directions. 
In order to construct exact density maps
we would need self-consistent models based on libraries of
non-periodic orbits (e.g. Schwarzschild 1979, Pfenniger 1984b,
Contopoulos \& Grosb{\o}l 1988).
This is beyond the scope of the present paper. Blurred
images, however, show, in a first approximation, the coarse
morphological features which are expected to be discernible in models
made out of non-periodic orbits, trapped around the stable periodic
ones. Thus our blurred images should not be considered as exact, but
just as guiding the eye, for morphological features to be sought in the
observations. We did not include in our blurred images planar orbits,
which would appear just as thick straight-line segments on the
equatorial plane. 

\begin{figure}
\epsfxsize=8.0cm \epsfbox{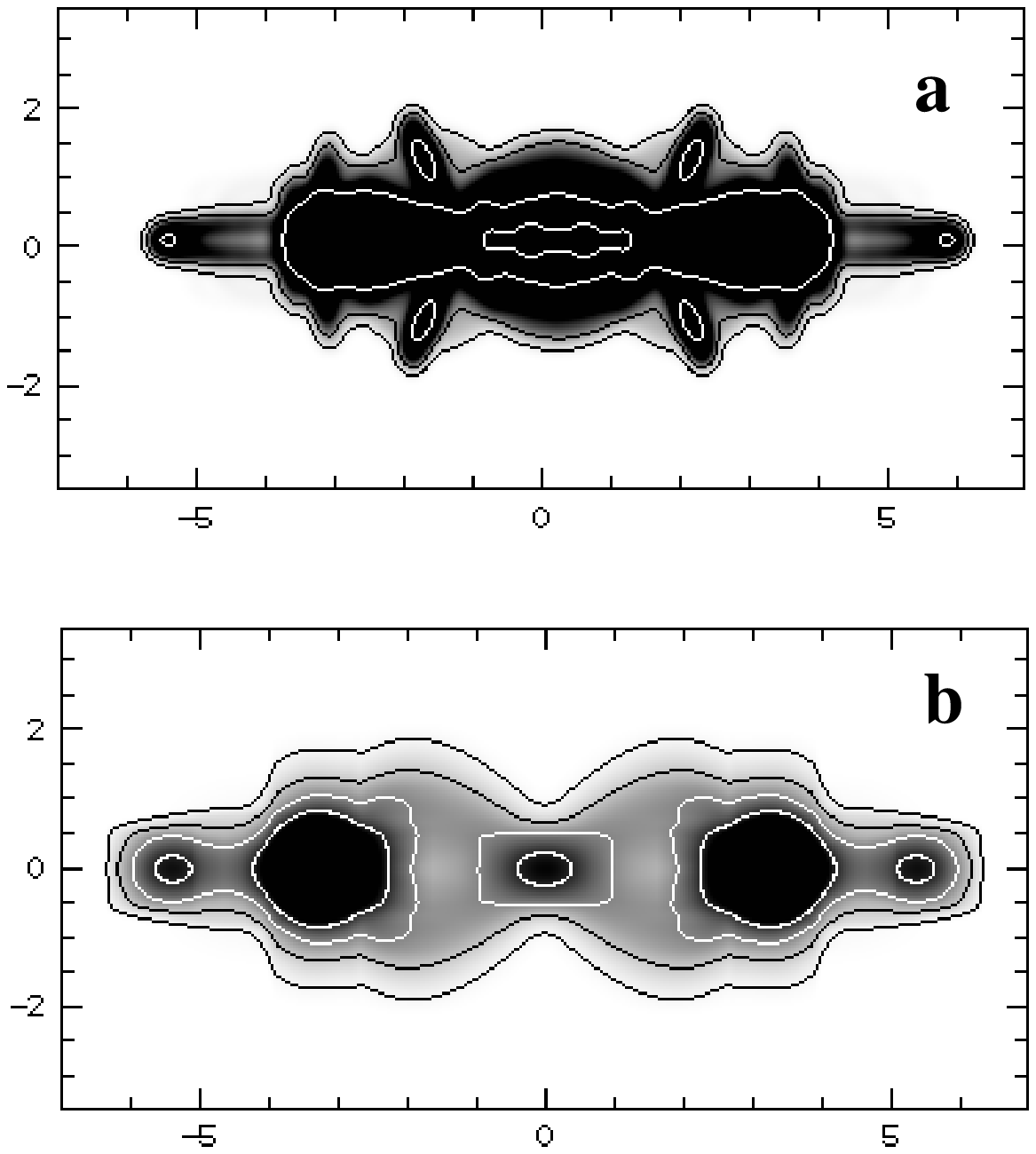}
\caption[]{Blurred images of the side-on profiles given in Fig.~\ref{Sc1yz}.
Characteristic isodensity contours indicate the morphologies which will be
favoured to appear in model A1.}
\label{BLSc1yz}
\end{figure}

The blurred images of the profiles of Fig.~\ref{Sc1yz} are given in
Fig.~\ref{BLSc1yz}. We clearly see that
we have boxy profiles both in (a) and in (b). The typical peanut morphology is
better reproduced in Fig.~\ref{BLSc1yz}b indicating that it is the x1v4 family
that is mainly responsible for this morphology. On the other hand, the boxy
structure in Fig.~\ref{BLSc1yz}a (as the overplotted isodensities also show)
is characterized by a local maximum height at $y$=0. We note that families
associated with vertical $n$:1 resonances with $n>4$ have little, if any,
contribution to the vertical structures that dominate and characterize the
profiles.

\begin{figure}
\epsfxsize=8.0cm \epsfbox{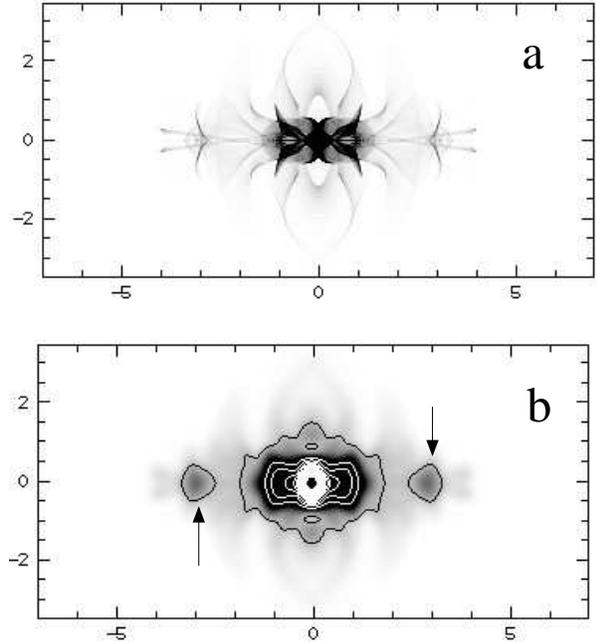}
\caption[]{Composite end-on, $(x,z)$, vertical profiles of model A1. (a) All 3D
families of Fig.~\ref{Sc1yz}a are considered. (b) The blurred image of
(a) (see text). The arrows point to the local density enhancements on either
side of the boxy bulge.}
\label{Sc1xz}
\end{figure}

Fig.~\ref{Sc1xz} is the end-on composite profile of model A1, considering all
the families, as in Fig.~\ref{Sc1yz}a. Its blurred image (Fig.~\ref{Sc1xz}b)
clearly shows that also the $(x,z)$ profile has a boxy character, as the
$(y,z)$ one. Now the box, as expected, is confined closer to the center.
Isodensity contours also in this case help
understand the relative importance of the various features.  The main
contribution to the central dark box in Fig.~\ref{Sc1xz}b comes from
the orbits of the \mbox{x2-like} 3D family of multiplicity 2
(Fig.~\ref{A1xzall}g). We note that in the corresponding side-on
profile the contribution of this family to the overall morphology is
minimal, due to the confined extent of their $(y,z)$ projections (they
are elongated along the minor axis of the bar).  Note the two surface
density enhancements, symmetric with respect to the center, indicated
by arrows.

\begin{figure}
\epsfxsize=7.0cm \epsfbox{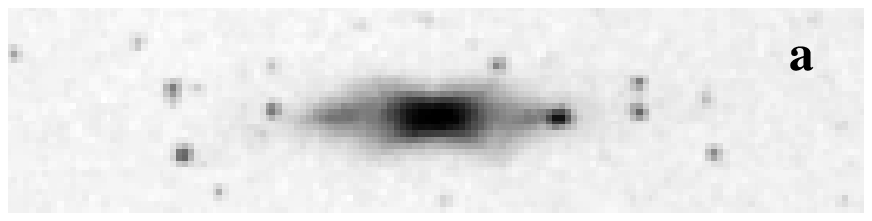}
\epsfxsize=7.0cm \epsfbox{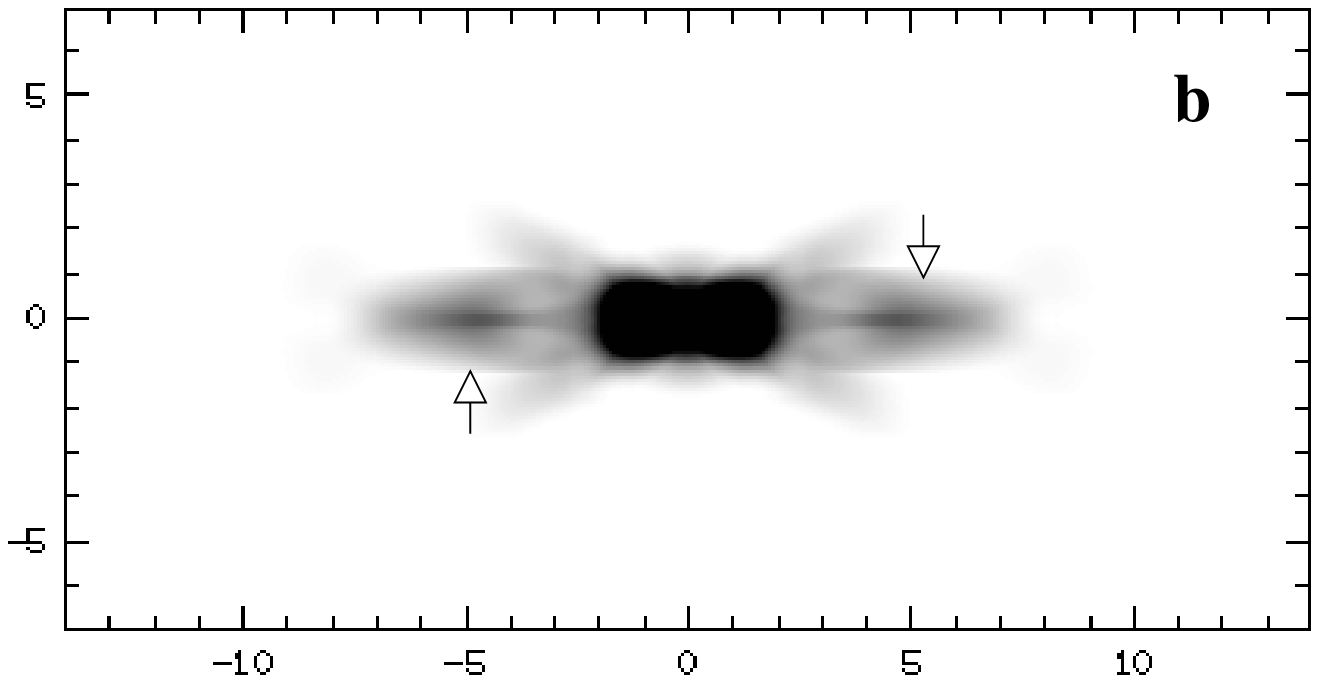}
\caption[]{(a) DSS image of IC~4767 in $B$. (b) the side-on, $(y,z)$,
  profile of model A2 made of the coexistence of the families x2v1, x1v1, x1v3
  and x1v4. The arrows point to the local density enhancements on either side
  of the boxy bulge.}
\label{ic_Sc2yz}
\end{figure}

Model A2, the slow rotating bar case, gives a characteristic example of a well
defined b/p shape due to successive orbits of the x1v1 family
(Fig.~\ref{A2yzall}b). Furthermore, it also gives a typical example of an {\sf
  `X'} morphology resulting from the superposition of successive orbits. The
main reason for the sharpness of the {\sf `X'} feature is that the S\ar \D\ar
S transition of the x1v1 family in this model practically does not interrupt
the succession of the stable orbits we consider, since x1v1 is complex
unstable only over a very narrow energy interval. In this way the {\sf `X'}
feature would be due to orbits trapped around the x1v1 family, rather than to
orbits trapped around families of stable orbits on inclined orbital planes
symmetric with respect to the rotational axis. The imprint of the {\sf `X'} in
model A2 is given in Fig.~\ref{ic_Sc2yz}b. Note
that the sides of the {\sf `X'} emerge out of a boxy feature in the very
center of the galaxy. Inspection of Fig.~\ref{A2yzall}b explains the origin of
this morphology. In the particular case of model A2, in the very inner part we
also have the contribution of family x2v1 with orbits with similar weights as
those of x1v1. Because of these, the {\sf `X'} feature does not extend with
equal intensity all the way to the center, unless the x2v1 are little
populated. Note that an {\sf `X'}
feature like the one presented by Whitmore \& Bell (1988) (their Fig.4) is
passing through the center and has equal intensity along its sides. Combining
the profiles of x1v1 and x2v1 with those of x1v3 and x1v4, we obtain an
edge-on profile of model A2 which has a striking similarity with the edge-on
galaxy IC~4767 (Fig.~\ref{ic_Sc2yz}a). We point with arrows at the
enhancements of the surface density along the major axis in
Fig.~\ref{ic_Sc2yz}b. They clearly have their counterparts in the image of the
galaxy (Fig.~\ref{ic_Sc2yz}a) and are clearly revealed in the processed image
in Whitmore \& Bell (1988, their Fig.~1c).

\begin{figure}
\epsfxsize=8.0cm \epsfbox{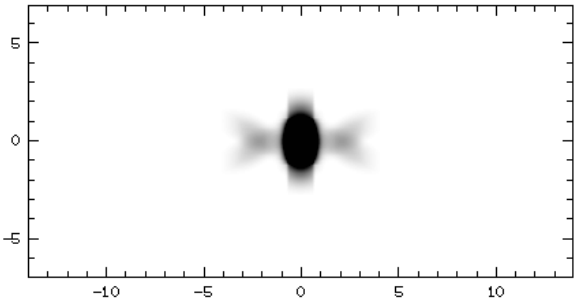}
\caption[]{Composite end-on, $(x,z)$, profile of model A2 given as a blurred
image. The profile is build by the same families used in Fig.~\ref{ic_Sc2yz}b.}
\label{Sc2xz}
\end{figure}

Let us now turn to the end-on view of model A2 (Fig.~\ref{Sc2xz}).  The family
x2v1 contributes a cross-type end-on profile. The remaining families
contribute mainly to the building of a dark bulge at the center, which is
slightly elongated along the rotational $z$ axis.
However, in this case corotation is at $r=13.24$, so we have, even by
considering x2v1, a ratio $R_{cor}/O_{Lx} \approx 3.6$, i.e. such features
should be sought in the very central regions of end-on views of the discs of
barred galaxies.

\begin{figure}
\rotate[r]{
\epsfxsize=8.0cm \epsfbox{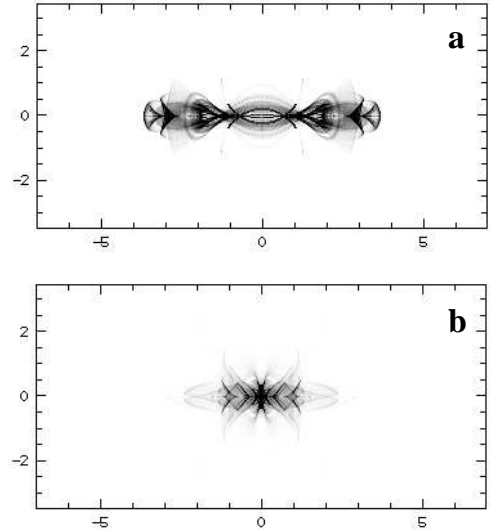}
}
\caption[]{Composite profiles of model A3. (a) $(y,z)$. (b) $(x,z)$.}
\label{Sc3prof}
\end{figure}

The fast rotating bar of model A3 offers an example, where the $(y,z)$
profile is rather flat (Fig.~\ref{Sc3prof}a), mainly due to the complex
instability of the
x1v1 family. The {\em end-on} profile is in this case boxy
(Fig.~\ref{Sc3prof}b).
The boxiness is introduced by the families x1v3 and x1v5.

\begin{figure*}
\begin{center}
\rotate[r]{\epsfxsize=10.0cm \epsfbox{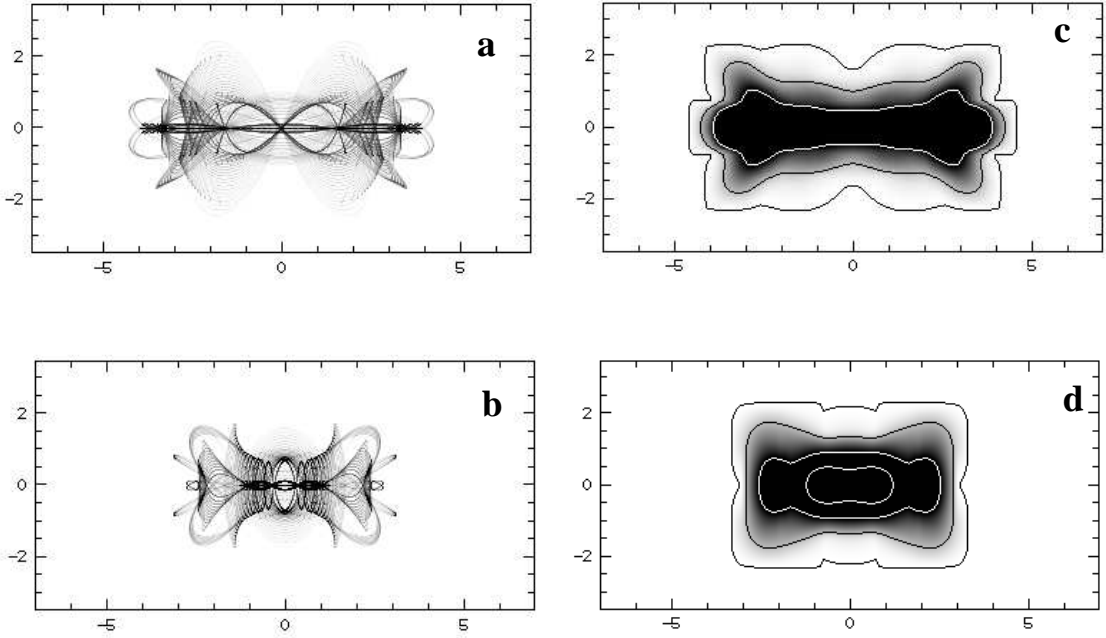} }
\end{center}
\vspace{-1cm}
\caption[]{Composite profiles of model B made out of x1v5, x1v5$^{\prime}$,
x1v7 and
z3.1s orbits. (a) $(y,z)$. (b) $(x,z)$.
In (c) and (d) are given the blurred images of (a) and (b), respectively.
Characteristic isodensity contours indicate the morphologies which will be
favoured to appear in model B.}
\label{SMprof}
\end{figure*}

Model B, without either radial or vertical 2:1 resonances, offers many
possibilities of showing a strong b/p feature if we consider all
families that may contribute to its vertical profile. It even has
a `\cx' feature in its central part when viewed side-on (Fig.~\ref{SMprof}a).
Note that in this case x1v1 orbits do not exist, since the first vertical
bifurcation in model B is x1v5 introduced in the vertical 4:1 resonance (paper
II). The profile is made out of x1v5, x1v5$^{\prime}$, x1v7 and z3.1s orbits,
which give the `\cx' morphology in the center. Blurred images are given in
Fig.~\ref{SMprof}c and Fig.~\ref{SMprof}d for the profiles in
Fig.~\ref{SMprof}a and Fig.~\ref{SMprof}b (end-on view), respectively. The
morphological 
difference between the x1v1 {\sf `X'} and the z3.1s {\sf `X'}, is that the
latter is characteristically curved (i.e. it can be better described with the
symbol `\cx'), while the sides of the former are straight. The overplotted
isophotes on the blurred images in Figs.~\ref{SMprof}c,d reveal a peanut and a
boxy profile respectively. From Figs.~\ref{Myz} and \ref{SMprof} it is
evident that the families x1v5,
x1v5$^{\prime}$ support a boxy structure elongated along the major axis with,
in the outer parts, an ansae-like morphology.
On the other hand the outer isophotes of z3.1s
support the typical peanut morphology, while closer to the equatorial plane
the isophotes have the kind of b/p shape encountered in several 3D bars of
$N$-body models (Athanassoula 2002, unpublished). 

\begin{figure}
\epsfxsize=8.0cm \epsfbox{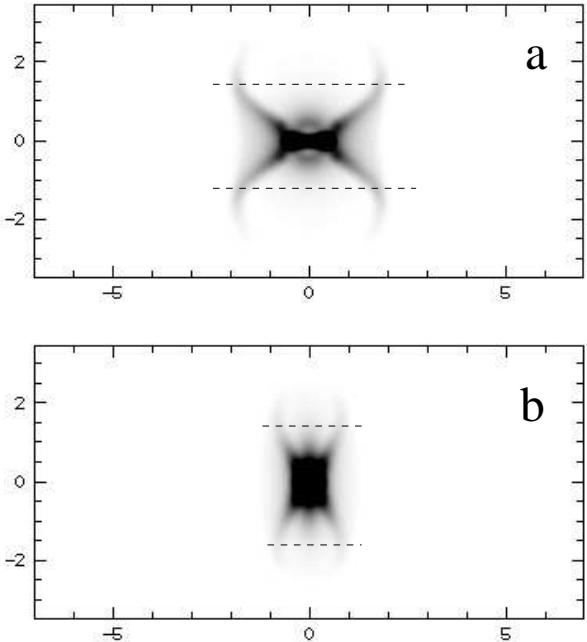}
\caption[]{Blurred representation of the side-on (a) and end-on (b)
  profiles of the x1v1 family in model C. The dashed lines indicate
  the region beyond which x1v1 orbits contribute to the local density only
  by orbits growing their sizes in $\overline{|z|}$ without increasing
  their projections along the major axis.  }
\label{Cxz}
\end{figure}

In Fig.~\ref{Cxz} we give in blurred representation also the edge-on (a)
and end-on (b) profiles of the x1v1 family in model C. In both figures
we have considered even orbits that reach $|z| > 1.5$, in order to show that
the x1v1 orbits do not contribute to the observed boxy feature beyond the
critical energy for which they start to increase their mean radii by increasing
practically only their mean $|z|$ values.  Beyond the dashed lines, away from
the equatorial plane, the corresponding density is too low. We see that a bar
with such a family as backbone, supports a peanut edge-on and a boxy end-on
profile, and also that we have here a case where the x1v1 family supports the
{\sf `X'} feature in its central morphology.

\begin{figure*}
\rotate[r]{
\epsfxsize=10.0cm \epsfbox{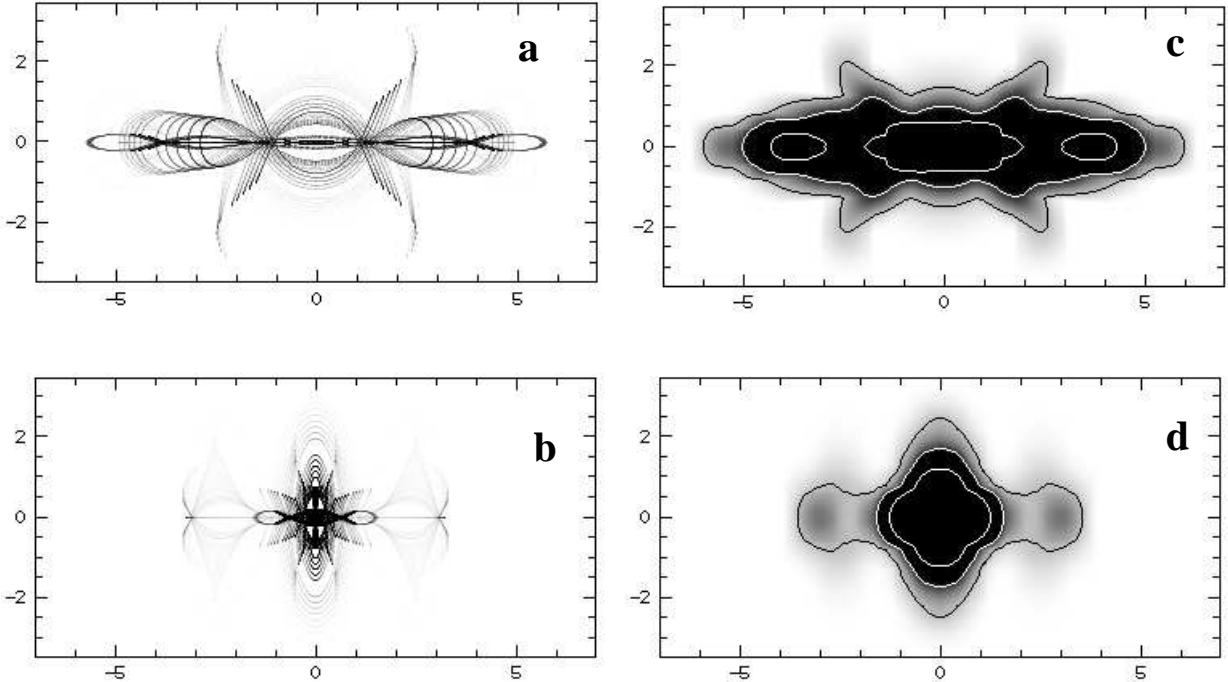}}
\caption[]{Composite profiles of model D with a strong bar. (a) $(y,z)$, (b)
  $(x,z)$, (c) blurred $(y,z)$ and (d) blurred  $(x,z)$.
  In panels (c) and (d) we include isocontours characteristic of the
  morphology of this model.
  It is evident that
  strong bars are associated with b/p side-on profiles and favour the presence
  of an {\sf `X'} feature.}
\label{SHprof}
\end{figure*}

Finally in model D (Fig.~\ref{SHprof}) we have another case characterized
by the dominance of x1v1. This family brings in the side-on morphology
a b/p bulge and broken branches of a {\sf `X'} feature. The end-on profile of
this model gives a rather roundish or rhomboidal central feature, which
becomes elongated in the $z$ direction if we take into account orbits with
mean $|z| > 1.5$ (Fig.~\ref{SHprof}d).

\section{Discussion}
In this paper we have investigated families of periodic orbits, as well as
combinations of such families, that might be the building blocks for b/p
edge-on morphologies in disc galaxies. The $B_L/O_{Ly}$ ratios in the side-on
views of the models vary. Profiles with $B_L/O_{Ly}$ ratios less than 2
characterize models in which most of the bar's material forms a b/p structure.
On the other hand, in cases with big $B_L/O_{Ly}$ ratios we have to do with
b/p features concentrated close to the center of the model. In almost all
cases (except for the z3.1s family) the orbits that contribute to a b/p
profile are introduced in the system as bifurcations of the basic family x1 at
vertical resonances. For this reason we used orbits of these families in
building possible orbital profiles.

In our study we present the orbital profiles as viewed from the two extreme
edge-on viewing angles, namely the side- and the end-on view. Possible changes
in the morphology of the profiles due to rotation have been discussed in
Section 3.1. There, we investigated the role of rotation in minimizing the
effect of the presence of the third local maximum in the profiles of the x1v1
family in model A1. We concluded, that the morphological change from a
structure with three local maxima to a kind of {\sf `X'}-feature happens only
for a small angle range. We find in general that the profiles keep their
morphological similarity with their side-on views for viewing angles close to
it, and the same happens with the end-on projections.  If we rotate a b/p
profile around the axis of rotation of the system, starting from the side-on
view, we observe that the local minimum at its center rises above the
equatorial plane and the profile becomes `thicker' at the center.  This,
however, does not affect the overall morphology of a b/p structure.  We note
that in blurred images of side-on projections, this minimum does not reach the
equatorial plane (Fig.~\ref{BLSc1yz}b, Fig.~\ref{ic_Sc2yz}b). Nevertheless,
the size of the bar altogether varies in different projections.

L"utticke et al. (2000b) hold the view that the inner regions of barred disc
galaxies, apart from the thin bar, are built by two further components,
namely a spheroidal bulge and a b/p structure. The spheroidal is a
kinematically distinct component occupying the central region of the galaxy.
Our models do not exclude the coexistence of spheroidals in the central
regions of the disks. Furthermore, we find families of periodic orbits in the
x1-tree, i.e. families of the 3D disc, that could support roundish structures,
especially in their end-on projections. It is also possible, as we found
z3.1s, to find a plethora of `z$n$' families that stay close to the rotational
axis and could populate a spheroidal bulge (e.g. family z5.1s given in Fig.~13
of paper II). These orbits are not presented in this paper, since we discuss
here only the presence and morphology of peanuts or boxes. Peanuts
and boxes are structures associated with families of the bar, or, more
generally, with families of the disc, since they can exist also in almost
axisymmetric models (Patsis et al. 2002a), as well as in the case where instead
of a bar we have a spiral perturbation (Patsis \& Grosb{\o}l 1996).  Note also
that the z3.1s family is bar-supporting (paper II).

Peanuts made out of the x1v1 family (i.e. related to the vertical 2:1
resonance) and in which also orbits beyond the \D\ar S transition are
populated show a spheroidal structure in the middle, which introduces a third
local maximum at the centers of the side-on profiles. The central empty region
of the `\emp' feature is surrounded by parts of x1v1 orbits. Branches of a
{\sf `X'} are stuck in several 
cases to the left and to the right of this spheroidal structure. We can
clearly see it in Fig.~\ref{Sc1yz}a, between the features pointed with arrows,
and also in the corresponding position of the edge-on profile of the strong
bar model D (Fig.~\ref{SHprof}a). In these cases the ratio of the extent of
the central pseudo-bulge over the extent of the peanut-shaped structure is
almost 1.

Geometrical arguments are also useful when comparing b/p features of real
galaxies or b/p features encountered in $N$-body simulations with b/p
structures in our orbital models. Unfortunately the various lengths involved
are measured in different ways by different authors. Usually the way of
measuring a length is intrinsic in the approach adopted by the authors in
order to study edge-on profiles. Thus a direct comparison is
problematic.

The peanut- or box-like features we find in the composite side-on profiles
have $B_L/O_{Ly}$ ratios which vary from about 2.5 to close to 1.  In our
measurements the length of the bar $B_L$ is the length of the longest
projection of bar-supporting orbits on the semi-major axis. In general these
are x1 orbits with loops on the major axis of the bar, or 2D rectangular-like
orbits at the radial 4:1 resonance region (Paper IV). Thus, an upper limit to
this length is defined in a rather precise way. On the other hand, the orbital
length along the major axis, $O_{Ly}$, can not exceed the length of the
projection of the outermost orbit of a family on the semi-major axis, 
but it can in principle take smaller values, depending on whether all 
or part of the family is
populated in the profile we construct. In that sense the $B_L/O_{Ly}$ ratios we
give can be considered as minimum values for a profile dominated by a specific
family, since we usually exclude only orbits that reach big heights above the
equatorial plane, which do not contribute much to density profiles anyway.
Nevertheless, if a family characterizes a profile this means that the majority
of its orbits is populated and in this case the ratio cannot be considerably
larger than the number we give.

By taking all the above into account we could compare the ratios we find with
corresponding quantities defined in Athanassoula \& Misiriotis (2002).
This paper examines quantitatively morphologies found in $N$-body
models of barred galaxies. In an
edge-on projection they estimate the extent of the bar and of the b/p feature
by considering cuts of the projected surface density parallel to the
equatorial plane. In the simulations where they find a strong b/p feature, the
ratio of the maximum distance to which the ledge on the $z=0$ cut extends to
the radius of very steep drops on cuts offset from $z=0$, is $\approx 1.5$.
This points to profiles dominated by orbits of families like x1v3, x1v4 or
x1v5, which in their side-on profiles have $B_L/O_{Ly}$ ratios less than 2.
Despite the differences that can be introduced due to the different kind of
modeling and the different ways of estimating the various relative lengths,
the b/p features in Athanassoula \& Misiriotis (2002) definitely occupy
regions larger than the central region of the galaxy. This indicates that a
large fraction of the bar seen side-on has a b/p morphology.

L"utticke et al. (2000b) find in their sample a ratio around 2.5 for the bar
length over the b/p distortion (BAL/BPL). Here again we have to note the
differences in the way they estimate the length of the bar and the length of
the b/p structure. As in Athanassoula \& Misiriotis (2002) they consider cuts
parallel to the equatorial plane of edge-on disc galaxies. The projected bar
length, however, is estimated in the cut along the major axis at the distance
where they find an increasing light distribution towards the center compared
to the radial exponential light distribution. The length of the b/p structure
is defined as the distance between the maxima of the b/p distortion.
Therefore, their bar length is systematically larger than both our $B_L$ and
the bar length estimated in Athanassoula \& Misiriotis (2002). Also BPL in
L"utticke et al. (2000b) is systematically smaller than our orbital length
$O_{Ly}$, since for most families parts of the orbits extend also to the left
and to the right of the maxima of the b/p feature. An exception is family x1v1
for which the distance between the b/p maxima and the length of the projection
of the orbital profile on the semi-major axis indeed almost coincide (see e.g.
Fig.~\ref{A1yzall}a). As a result of all the above mentioned differences the
BAL/BPL ratio in L"utticke et al. (2000b) is expected to be systematically
larger than our $B_L/O_{Ly}$ in profiles dominated by the same family.  For
our models ratios around 2.5 point to x1v1 type peanuts or, in models without
radial and vertical ILRs, to peanuts that are made of z3.1s orbits. This,
however, does not exclude that many galaxies in the L"utticke et al. (2000b)
sample could have $B_L/O_{Ly}$ less than 2 for the family which is mainly
responsible for its vertical profile.

As a general rule we can say that, if the relative size of a b/p feature
compared with the size of the bar is estimated smaller than 2, this indicates
that most probable candidates to be associated with it are families related to
higher order vertical resonances (e.g. x1v3, x1v4 etc.) and not the x1v1
family. The main reason for this is that the x1v1 orbits, as energy increases
approaching corotation, do not increase their projections on the major axis of
the bar after a critical energy. Beyond this energy, the x1v1 orbits grow
practically only in $z$.

For cases of edge-on galaxies where the ratio of the bar to the b/p
distortion's length is larger than 3, the orbital models indicate that
these features may be associated with nearly end-on projections of
various families (see Fig.~\ref{A1xzall}b and g, Fig.~\ref{A3xzall}b
and c, Fig.~\ref{Hxz}b). Also existing x2-like 3D orbits would
contribute to boxy end-on profiles. x1v1, the family associated with
the vertical 2:1 resonance, does not give boxy end-on projections. It
contributes rather to a roundish central morphology
(Figs.~\ref{A1yzall}a, \ref{Hyz}a, \ref{SHprof}d). At this point we
want to stress that side-on b/p profiles combined with slightly
prolate end-on morphologies, as well as boxy, or even b/p, end-on
profiles are encountered in 3D bars in $N$-body simulations
(Athanassoula 2002, unpublished).

Another feature that we can geometrically quantify is the oblique angle
between a branch of {\sf `X'} and the major axis of the main bar. In the slow
bar model A2 the {\sf `X'} is made mainly of the x1v1 orbits and this angle is
about $27\degr$ (Fig.~\ref{A2yzall}b), close to the value Whitmore \& Bell
(1988) give for IC~4767. The x1v1 orbits of models A1 (fiducial case) and D
(strong bar case) harbour {\sf `X'}s embedded in peanut-shaped bulges, which
are characterized by the fact that they do not reach the major axis. As we
explained, this results from the presence of the complex unstable part of this
family which causes an empty inner region. In these cases the angle is about
$50\degr$.
In the strong bar case (model D) the peanut feature with an {\sf `X'} is
particularly conspicuous, in agreement with the results of $N$-body
simulations (Athanassoula \& Misiriotis 2002, and Athanassoula 2002,
unpublished).
The central `\cx' feature made out of the z3.1s orbits in the model
without 2:1 resonances does not include segments with straight lines. However,
tangents crossing through the center are inclined about $43\degr$ to the major
axis. We can define an inner and an outer inclination for branches of the
`\cx' features due to the x1v4 family that might appear in several models. In
model A1 we measure this to be about $25\degr$ if we consider stable x1v4
orbits with low energy, and $50\degr$ if we consider stable x1v4 orbits with
large energies (Fig.~\ref{Sc1yz}b). A rough estimation of the angle of {\sf
  `X'} in Hickson 87a from Fig.~4 in Mihos, Walker and Hernquist (1995), is
about $30\degr$. L"utticke et al. (2000c) estimate this angle to be $40\degr
\pm 10$. To summarize, small `{\sf `X'}-angles' ($< 30\degr$) are build from
orbits related to families introduced at higher order vertical resonances and
to some degree slow rotating bars, while large `{\sf `X'}-angles' ($>
40\degr$) characterize {\sf `X'} structures in typical x1v1 b/p bulges.

The overlapping of periodic orbits of some families form two enhancements of
the projected surface density along the major axis of the models, which 
are symmetric
with respect to the center. Similar features, seen along the major axes of
edge-on disc galaxies with b/p bulges, e.g. IC 4767 \cite{whib88}, ESO 417-G08,
NGC 4710 \cite{schd00} are often interpreted as rings or spirals. Here we
present an alternative explanation. Using families of periodic orbits as
building blocks of the profiles of the edge-on barred galaxies we can easily
see that the enhancements could be a kind of a projection effect due to the
trapping of material around stable periodic orbits. These enhancements
actually exist in the profiles of all families. In composite profiles one can
better see them at families with orbits close to the end of the bar. The
reason is that there are no other families of periodic orbits that fall on
them smoothing out the profile more. Recently Aronica, Bureau,
Athanassoula et al. (2002, and in preparation), detected these enhancements 
in a sample of edge-on
galaxies observed in the K$^{\prime}$.

We note that we found b/p features in models without a radial 2:1 resonance
(model C, cf. Fig.~\ref{Cxz}) and that we encountered a case (model A3) where
although the radial and vertical 2:1 resonances exist and are located
at very close energy values (paper II), the model does not support a
conspicuous b/p structure
(cf. Fig.~\ref{A3yzall}). Thus in our response models we do not find any
correlation of a conjuction of radial and vertical 2:1 resonance and the
appearance of a b/p feature.

Finally our models could support small inner discs, tilted with respect to the
equatorial plane of the main disc. This can be done by breaking the symmetry
and considering only one of the symmetric branches of a family. In galaxies
this can be e.g. due to companions falling on a target galaxy at a skew angle,
so that the system will show a preference in populating one of the two
symmetric branches.

\section{Conclusions}
In this paper we investigated the vertical structure in barred potentials
using orbital
theory. We combined families of periodic orbits in the models of papers I and
II in order to study the possible resulting vertical profiles and we discussed
their geometry and their dimensions in comparison with structures found in
edge-on disc galaxies and snapshots of $N$-body simulations. Here we enumerate
our basic
conclusions:
\begin{enumerate}
\item The vertical profiles of our models are of `stair-type'. This
means that families that offer the skeletons for the 3D bars and are 
bifurcated at higher energies (i.e. closer to corotation) have 
  in general lower mean heights.
\item b/p features in vertical profiles can be supported mainly by the
  following families:
\begin{itemize}
\item {\bf x1v1}. This family is particularly useful for building a boxy
  central structure
  if it does not have a complex unstable part in the critical energy.
  The best
  examples we found are in the slow rotating bar case and in the strong bar
  model. In these cases, for energies where the maximum $z$ of the orbits
  remains less than about 1~kpc, successive orbits of x1v1 have the maximal
  deviations of their edge-on projections from the equatorial plane aligned
  along almost straight segments. These `lines' are in oblique angles to the
  major axis, not passing thorough the center in general, and their angle with
  the major axis is $\approx 50\degr$. In the slow bar case, however, this
  angle is $\approx 27\degr$. For this family the ratio {\em $B_L/O_{Ly}$ is
  larger than 2}. $B_L/O_{Ly}>3$ for x1v1 (as found in model A3), indicate
  that only a small part of the family is populated and thus its contribution
  to the vertical structure of the model is not significant.
\item {\bf x1v4}. This family gives $B_L/O_{Ly} \approx 1.3$, i.e. brings the
  end of the peanuts close to the end of the bars. It is introduced in the
  system after a U\ar S transition of x1 and has stable representatives for
  larger energies than the energy at the bifurcating point. It exists over
  large energy intervals and, if populated, will provide the system with a
  b/p-shaped structure whose extent is near that of the bar. It
  supports the peanut morphology especial in composite profiles
  without any contribution from x1v1 orbits (Fig.~\ref{BLSc1yz}b).
\item {\bf z3.1s} This family gives b/p features in models without radial or
  vertical ILRs, and is quite important for the dynamics of these
  models. Profiles characterized by the presence of this family
  (Fig.~\ref{SMprof}c, Fig.~\ref{Myz}c) have the characteristic local minimum
  of the density at the (0,0) position, like in the case of NGC~2788~A
  in Fig.~\ref{twogal}a, which is
  indicated by a black arrow.
\item The ansae-type profile is easiest made by orbits associated with the
  vertical 4:1 resonance and can be described as a stretched {\sf `X'}
  (Fig.~\ref{Myz}a).
\item Finally the 3D {\bf x2-like} families of orbits support boxy
  morphologies. The latter are especially discernible in the end-on profiles
  of the model with the low pattern speed.
\end{itemize}
We have to note that if families bifurcated at high order vertical resonances
are responsible for the peanut, then they will support a b/p bar morphology
altogether.  Coexistence of several 3D families should in general be expected.
In such a case it is the family which is bifurcated at the lowest energy 
that plays the most important role for the morphology of the model.

\item Narrow extensions appear on the sides of many profiles (see
  e.g. Fig.~\ref{BLSc1yz}a) These features result from the
  `stair-type' character of profiles constructed with families of
  periodic orbits, and have their counterpart in many images of
  edge-on disc galaxies. The corresponding feature in the case of
  NGC~6771 is also indicated by white arrows in Fig.~\ref{twogal}b.
\begin{figure}
\hspace{1.2cm}
\epsfxsize=5.0cm \epsfbox{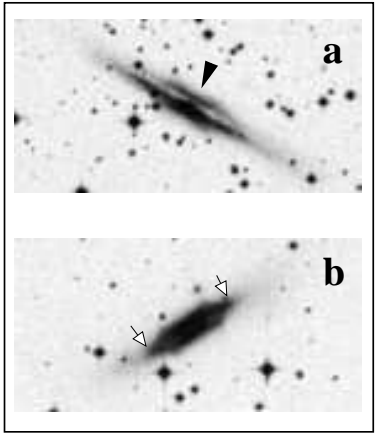}
\caption[]{DSS images of NGC~2788~A (a), and NGC~6771 (b). Both
  galaxies show a b/p profile.}
\label{twogal}
\end{figure}
\item The projection of the orbits of a family on the equatorial plane is
  confined within certain limits. By this we mean that moving on the
  characteristic towards corotation, we reach a certain distance from the
  center where the mean radius of the orbits increases only due to increasing
  of $z$. This is particularly evident in the case of the x1v1 family, which
  is related to the vertical 2:1 resonance and which in general is the 3D
  bifurcation of x1 closest to the center.
\item Families of periodic orbits (x1v3, x1v4, x1v5, z3.1s, x2v1) can build
  boxy, or even peanut-boxy, {\em
  end-on} profiles. We would thus like to suggest that boxy bulges in 
  galaxies having a bar
  length over a b/p length larger than 3, are related with the
  profiles of families seen end-on. 
  \item {\sf `X'}-type features are found in the composite orbital profiles.
  They are formed by alignment of successive orbits of the family x1v1.  They
  are pronounced if the S\ar \De transition in this family does not play an
  important role in the dynamics of a model. The fact that {\sf `X'} features
  are rare in real galaxies, indicates that, in most cases where a x1v1 
  family is populated in
  a galaxy, it has an important complex unstable part.  Adequate successive
  projections of the orbits of a family in large energy intervals, like what
  we see in model B mainly due to the z3.1s orbits, give central morphologies
  that can be described with the symbol `\cx'. We note that the higher the
  order of the resonance of the family associated with {\sf `X'} or `\cxs'
  structure is, the smaller the angle between the {\sf `X'}/`\cx'-branch and
  the major axis we find. In the case of the x1v1 family this angle is smaller
  in the slower rotating bar case.
\item Discs out of the equatorial plane in the bulges are easiest made by
  breaking the symmetry and populating only one of the two symmetric branches
  of the 3D families.
\item Characteristic local enhancements of the surface density along the major
  axis of the bar are predicted by the models merely due to the orientation of
  the successive orbits in the profiles.
\end{enumerate}

\section*{Acknowledgments}

We acknowledge fruitful discussions and very useful comments by G.~Contopoulos
and A.~Bosma. We thank the referee for useful suggestions and valuable
remarks, which improved the paper.
This work has been supported by E$\Pi$ET II and K$\Pi\Sigma$
1994-1999; and by the Research Committee of the Academy of Athens.  ChS
and PAP thank the Laboratoire d'Astrophysique de Marseille for an
invitation during which essential parts of this work have been completed.
ChS was partially supported by the ``Karatheodory'' postdoctoral
fellowship No 2794 of the University of Patras. All image processing work has
been done with ESO-MIDAS.

\bsp

\label{lastpage}

\end{document}